\documentclass[a4paper,12pt,aps,superscriptaddress]{article}
\usepackage{amsmath}
\usepackage{graphicx}
\usepackage{epsf}

\renewcommand{\phi}{\ensuremath{\varphi}}

\begin{document}

\title{Canonical Analysis of Condensation in Factorised Steady States}
\date{\today }
\author{M.\ R.\ Evans$^1$, Satya\ N.\ Majumdar$^{2}$, R.\ K.\ P.\ Zia$^{3}$ \\
%EndAName
$^1$SUPA, School of Physics, University of Edinburgh,\\
Mayfield Road, Edinburgh EH9 3JZ, UK\\
[0.5ex] $^2$ Laboratoire de Physique Th\'eorique et Mod\`eles Statistiques,\\
Universit\'e Paris-Sud, Bat 100, 91405, Orsay-Cedex, France\\
[0,5ex] $^3$Department of Physics and\\
Center for Stochastic Processes in Science and Engineering,\\
Virginia Tech, Blacksburg, VA 24061-0435, USA}
\maketitle

\begin{abstract}
We study the phenomenon of real space condensation in the steady state of a
class of mass transport models where the steady state factorises. The grand
canonical ensemble may be used to derive the criterion for the occurrence of
a condensation transition but does not shed light on the nature of the
condensate. Here, within the canonical ensemble, we analyse the condensation
transition and the structure of the condensate, determining the precise
shape and the size of the condensate in the condensed phase. We find two
distinct condensate regimes: one where the condensate is gaussian
distributed and the particle number fluctuations scale normally as $L^{1/2}$
where $L$ is the system size, and a second regime where the particle number
fluctuations become anomalously large and the condensate peak is
non-gaussian. Our results are asymptotically exact and can also be
interpreted within the framework of sums of random variables. We further
analyse two additional cases: one where the condensation transition is
somewhat different from the usual second order phase transition and one
where there is no true condensation transition but instead a
pseudocondensate appears at superextensive densities.

\noindent

\end{abstract}

\section{Introduction}

Mass transport models encompass a large class of systems wherein `mass', or
some conserved quantity, is transferred stochastically from site to site of
a lattice. Well known examples of such models are the Zero-Range Process
(ZRP) \cite{MRE00}, and the Asymmetric Random Average Process (ARAP) \cite
{ARAP}. These are simple fundamental models which have been used to describe
such diverse physical situations as traffic flow \cite{traffic}, clustering
of buses \cite{OEC}, phase separation dynamics \cite{KLMST}, force
propagation through granular media \cite{CLMNW}, shaken granular gases \cite
{MWL04,Torok} and sandpile dynamics \cite{Jain}. For a recent review of the
ZRP and related models and applications see \cite{EH05}.

Condensation is manifested in such models when, in the steady state, a
finite fraction of the total mass condenses onto a single lattice site. At
low global mass densities the system is in the fluid phase where the mass is
distributed evenly over all sites. Here, the single site mass distribution $%
p(m)$ typically decays exponentially with $m$ implying a finite amount of
mass at each site with the mean being the global mass density $\rho $. On
increasing the global density above a critical value, $\rho _c$, an extra
piece of $p(m)$ emerges which represents the condensate i.e. a single site
which contains the global excess mass $(\rho -\rho _c)L$ where $L$ is the
system size. Thus in the condensed phase the condensate coexists with the
background fluid. Note that the condensation occurs in real space and, in
principle, in any number of dimensions. In this paper we will analyse in
detail the structure of the condensate, elucidating when it occurs, its
shape and its fluctuations. A short communication of some of our results has
been given in \cite{MEZ05}.

In general mass transfer models form examples of systems exhibiting
nontrivial nonequilibrium steady states. Since the models are defined by
stochastic dynamical rules, their steady state distributions are not a
priori known. However it turns out that a large class of models, which we
shall discuss below, enjoy the convenient property of having a factorised
steady state. This means that the steady state probability $P(\{m_l\})$ of
finding the system with mass $m_1$ at site 1, mass $m_2$ at site 2 etc is
given by a product of (scalar) factors $f(m_l)$ --- one factor for each site
of the system --- i.e. 
\begin{equation}
P(\{m_l\})=Z(M,L)^{-1}\prod_{l=1}^Lf(m_l)\;\,\delta \!\left(
\sum_{l=1}^Lm_l-M\right) \;.  \label{prob}
\end{equation}
where $Z(M,L)$ is a normalisation which ensures that the integral of the
probability distribution over all configurations containing total mass $M$
is unity, hence 
\begin{equation}
Z(M,L)=\prod_{l=1}^L\left[ \int_0^\infty \ensuremath{\mathrm{d}}%
m_lf(m_l)\right] \,\delta \!\left( \sum_{l=1}^Lm_l-M\right) \;.  \label{Z}
\end{equation}
Here, the $\delta $-function has been introduced to guarantee that we only
include those configurations containing mass $M$ in the integral. The
single-site weights, $f(m)$ are determined by the details of the mass
transfer rules.

Thus when the steady factorises, the analysis of condensation reduces to the
evaluation of (\ref{Z}), a problem first addressed in \cite{BBJ}. Equation (%
\ref{Z}) defines a canonical partition function in that the delta function
imposes the constraint of fixed total mass $M$. As we shall discuss in
detail below analysis of (\ref{Z}) has only been carried out in the fluid
phase and effectively in the grand canonical ensemble where the total mass
is allowed to fluctuate. Our aim here is to present a full analysis of (\ref
{prob},\ref{Z}) in the canonical ensemble. In particular this shall
elucidate the mechanism of condensation within factorised steady state.

Although our focus in this paper is on the analysis of the steady state (\ref
{prob},\ref{Z}), it is relevant to briefly review some particular models and
dynamics which give rise to such steady states, and previous studies of
condensation phenomenon. Firstly we mention the backgammon model \cite
{Ritort} where unit masses hop under dynamics respecting detailed balance
with respect to an energy function which is simply minus the number of
unoccupied sites in the system. When the temperature $T \to 0$ a condensate,
where all masses are on the same site, dominates the steady state. Since the
model and dynamics are constructed so that an energy functions exists, the
steady state automatically factorises into the form (\ref{prob},\ref{Z})
(but with discrete mass variables). The motivation for the Backgammon Model
was actually to study a simple model for glassy dynamics since in the late
time regime at low $T$ entropic barriers and slow dynamics arise.

In \cite{BBJ} the factorised steady state of the Backgammon model was
generalised to the form (\ref{prob},\ref{Z}) with single-site weights $f(m)$
that could have an asymptotic power law dependence 
\begin{equation}
f(m)\sim m^{-\gamma} \;.  \label{f(m)}
\end{equation}
It was shown that a condensation transition would occur if $\gamma >2$.
Mean-field dynamics for this general factorised steady state were
constructed from the detailed balance condition and the mean-field dynamics
of condensation was studied in \cite{DGC}. Here `mean field' is used in the
sense that a particle can hop from a given site to any other.

On the other hand it has long been known that the ZRP, a stochastic model
defined by \emph{local} stochastic dynamics and without respect to any
energy function, has a factorised steady state \cite{S70,MRE00}. In a
heterogeneous system, where the rules for mass transfer depend on the site,
the condensation may occur at the site with lowest outgoing mass transfer
rate \cite{E96,KF96}. In this case the mechanism for condensation (in real
space) is exactly analogous to Bose-Einstein condensation in momentum space
in an ideal Bose gas and the slowest site plays the role of the ground state
in the quantum system.

For homogeneous dynamical rules for the transfer of mass, which we will
focus on in this work, the condensation mechanism in the ZRP was first
studied, to our knowledge, in \cite{OEC} where the ZRP was used as an
approximate description of a model of bus routes. It was shown that if the
hop rate $u(m)$ for a mass to move from a site with $m$ masses to its
neighbouring site decays as $u(m) \sim \beta(1+ \gamma/m)$ then the
single-site weight decays as (\ref{f(m)}) so that a condensation transition
occurs if $\gamma >2$. Condensation in the ZRP has been further studied in 
\cite{JMP00,GSS,Godreche,AEM}.

Another model with a factorised steasy state is the ARAP \cite{ARAP}.
Although usually this does model does not exhibit condensation, condensation
may be induced by imposing a maximum threshold on the amount of mass that
may be transferred between sites \cite{ZS2}.

It turns out that the ZRP and ARAP may be unified by considering a very
general class of one-dimensional mass transport models\cite{EMZ04}: a mass $%
m_i$ resides at each site $i$; at each time step, a portion, $\tilde{m}_i\le
m_i$, chosen from a distribution $\phi (\tilde{m}|m)$, is chipped off to
site $i+1$. Choosing the chipping kernel $\phi (\tilde{m}|m)$ appropriately
recovers the ZRP, the ARAP and the chipping model of~\cite{MKB}. Moreover
the class of models encompasses both discrete and continuous time dynamics
and discrete and continuous mass. A necessary and sufficient condition for a
factorised steady state is that the chipping kernel is of the form\cite
{EMZ04} 
\begin{equation}
\phi (\tilde{m}|m)\propto u(\tilde{m})v(m-\tilde{m})
\end{equation}
where $u(z)$ and $v(z)$ are arbitrary non-negative functions. Then the
single site weight is given by 
\begin{equation}
f(m)=\int_0^md{\tilde{m}}u(\tilde{m})v(m-\tilde{m})  \label{ffact}
\end{equation}
This result may be generalised to arbitrary lattices in all dimensions \cite
{EMZtbp,GL05}. Given a chipping kernel $\phi (\tilde{m}|m)$, sometimes it is hard
to verify explicitly that it is of the form (\ref{ffact}) and thereby to
identify the functions $u(m)$ and $v(m)$ in order to construct the weight $%
f(m)$. This problem was circumvented by devising a test~\cite{ZEM04} to
check if a given explicit $\phi (\tilde{m}|m)$ satisfies the condition (\ref
{ffact}) or not. Further, if it ``passes this test,'' the weight $f(m)$ can
be found explicitly~by a simple quadrature \cite{ZEM04}. Finally, for any
desired function $f(m)$, one can construct dynamical rules (i.e., $\phi $'s)
that will yield $f(m)$ in a factorised steady state.

Condensation has also been noted and studied in mass transport models
without factorised steady states. In the chipping model defined in \cite{MKB}
all the mass at a site can move to a neighbouring site. Condensation in the
model was studied in a mean-field approximation for the steady state which
amounts to approximating the true (unknown) steady state by the factorised
form (\ref{prob},\ref{Z}). It turned out that in the true steady state
condensation is actually only observed in the case of symmetric hopping \cite
{MKB,RM2,RK}. However a generalisation which includes both the chipping
model and ZRP as special cases does appear to exhibit condensation for
asymmetric hopping \cite{LMZ}. Finally we mention more complicated
condensation scenarios such as the two species ZRP \cite{EH03} where, for
example, one species may induce condensation in the other.

So far we have reviewed models exhibiting condensation which comprise
hopping particles. Condensation phenomena have also been observed in the
context of the rewiring dynamics of networks with a fixed number of nodes
and links \cite{DMS03,BCK01}. There the condensate corresponds to a node
which is `hub' which captures a finite fraction of the links. Also, models
of macroeconomies may exhibit `wealth condensation' where the wealth is the
conserved quantity and condensation corresponds to a single individual
capturing a finite fraction of the wealth\cite{BJJKNPZ02}. A further
important application of the condensation mechanism is to determine a
criterion for phase separation in one-dimensional systems \cite{KLMST,ELMM}.
Phase separation, or indeed the lack of it, has been reported in a number of
one-dimensional and ``quasi-1-D'' systems typically involving three species
of charged particles with an exclusion interaction \cite
{AHR,KSZ,RSS,MSZ,KLMT,KLMSW,GSZ}. By mapping the domains between the neutral
particles onto the sites of a ZRP, the ZRP can be used as an effective
description of domain dynamics and the criterion for condensation in this
effective description determines when phase separation will or will not
occur.

The paper is structured as follows. In section~\ref{sec:gce} we review the
grand canonical approach which correctly predicts the critical density, but
does not shed light on the nature of the condensate. Before embarking on the
canonical analysis we interpret in section~\ref{sec:rv} the canonical
partition function in the framework of sums of random variables and deduce
rather quickly some key properties of the condensate. In section~\ref
{sec:nat} we summarise the results of the full canonical analysis which are
detailed in sections~\ref{sec:exact} and \ref{sec:cangen}: in section~\ref
{sec:exact} we present an exactly solvable case where a closed expression
can be found for the canonical partition function and in section~\ref
{sec:cangen} the asymptotic analysis of the canonical partition function in
the general case is presented. In section~\ref{sec:oc} we consider other
forms of $f(m)$ such as (\ref{f(m)}) with $\gamma <2$ for which condensation
does not occur, and stretched exponential $f(m)$ for which condensation does
occur but with some differences. We conclude with a discussion in section~%
\ref{sec:disc}.

\section{Criterion for Condensation: Grand Canonical Approach}

\label{sec:gce} Assuming that the steady state of mass transport model is of
the factorised form (\ref{prob},\ref{Z}), one next turns to the issue of
condensation within this factorised steady state. In particular, we ask: (i)
when does a condensation transition occur, i.e., the criterion for
condensation (ii) if condensation occurs, what is the precise nature of the
condensate? The factorization property allows (i) to be addressed rather
easily within a grand canonical ensemble (GCE) framework -- an approach
similar to the one used in traditional Bose-Einstein condensation in an
ideal gas of bosons, though there are some important differences in the
present case. In this approach one takes the thermodynaimc limit ($L,M\to
\infty $ with fixed \emph{overall }density $\rho =M/L$). Then (\ref{prob})
becomes a product measure state where the sites are decoupled from each
other and the single site mass distribution is simply given by $p(m)=f(m)%
\mathit{e}^{-\mu m}$ where $\mu $ is the negative of the chemical potential
and is chosen to fix the density, giving rise to the relation 
\begin{equation}
\rho =\rho (\mu )\equiv \frac{\int_0^\infty mf(m)e^{-\mu m}%
\ensuremath{\mathrm{d}}m}{\int_0^\infty f(m)e^{-\mu m}\ensuremath{\mathrm{d}}%
m}.  \label{cp1}
\end{equation}
The criterion for condensation can be derived easily by analysing the
function $\rho (\mu )$ defined in Eq. (\ref{cp1}).

The behavior of $\rho(\mu)$ depends on $f(m)$. We consider three cases
separately.

\begin{enumerate}
\item  First consider the case when $f(m)$ decays faster than exponential
for large $m$. In this case, $\mu $ is allowed to take any value in $%
[-\infty ,\infty ]$ and $\rho (\mu )$ is monotonic in $\mu $, ranging from $%
\infty $ to $0$. Thus, for a given $\rho $, one can always find a suitable $%
\mu $ to satisfy Eq. (\ref{cp1}) and there is no condensation.

\item  We next consider the case when $f(m)$ decays slower than $m^{-2}$ for
large $m$. In this case, the allowed range of $\mu $ is $[0,\infty ]$, in
order that the integrals in (\ref{cp1}) converge. The function $\rho (\mu )$
diverges as $\mu \to 0$, and then decreases monotonically to $\rho (\mu )\to
0$ as $\mu \to \infty $. Once again, for any given $\rho $, there is always
a solution of $\mu $ from Eq. (\ref{cp1}) and there is no condensation.

\item  Finally consider the case when $f(m)$ decays (a) slower than
exponential but (b) faster than $m^{-2}$. In this case, the allowed range of 
$\mu $ is $[0,\infty ]$ over which $\rho (\mu )$ is a monotonically
decreasing function of $\mu $. However, the crucial difference here from the
two previous cases is that $\rho (0)$ is finite. Indeed, $\rho (0)$ sets the
critical density $\rho _c=\rho (0)$. When the given $\rho <\rho _c$, a
positive $\mu $ satisfying (\ref{cp1}) exists. However, when $\rho >\rho _c$%
, there is no real solution to (\ref{cp1}), signalling a condensation
transition. In this phase, the extra mass $(\rho -\rho _c)L$ forms a
condensate.
\end{enumerate}

A natural choice of $f(m)$ satisfying the criterion of condensation in 3.
above is 
\begin{equation}
f(m)\simeq A\,m^{-\gamma }\quad \mathrm{with}\quad \gamma >2.  \label{fm1}
\end{equation}
For the majority of this paper, we stay with the choice of $f(m)$'s in (\ref
{fm1}) and set, without loss of generality, $\int_0^\infty f(m)%
\ensuremath{\mathrm{d}} m=1$. For brevity, we also assume that $\gamma $ is
non-integer although our analysis can be easily extended to the integer
case. In section~\ref{sec:oc} some other forms for $f(m)$ will be explored.

The single-site probability distribution $p(m)$ is given by 
\begin{equation}
p(m) = f(m) e^{-\mu m} \quad\mbox{for}\quad \rho < \rho_c \;.
\end{equation}
which exhibits a characteristic mass $m^*= 1/\mu$. As the critical density
is approached $\mu$ decreases to zero and the characteristic mass diverges
giving rise to a power law $p(m) \simeq f(m)$ at the critical density. For $%
\rho> \rho_c$ the grand canonical approach can not be used to determine $%
p(m) $.

This last point is in contrast to usual Bose-Einstein condensation where one
can work in the GCE even in the condensed phase. This is done by letting the 
$\mu$ tend to zero as $1/V$ where $V$ is the volume of the system i.e. any
density of bosons can be achieved by carefully letting $\mu \searrow 0$ in a
way dependent on system size. However, in the present case letting $\mu \to 0
$ in (\ref{cp1}) always results in a fixed critical density $\rho_c$.

%{\bf added discussion of form of p(m)}

\section{Interpretation as sums of random variables}

\label{sec:rv} Before proceeding with a detailed analysis of the canonical
partition function (\ref{Z}) it is instructive to discuss the problem from
the perspective of sums of random variables.

First note that if $f(m)$ is normalised then $\prod_{i=1}^L f(m_i)$ is the
probability that $L$ independent and identically distributed (iid) positive
random variables, each drawn from a distribution $f(m)$, take the values $%
m_1,m_2,\ldots m_L$ which we denote by $\underline m$. Moreover, the
partition function $Z(M,L)$ in (\ref{Z}) is precisely the probability that
the sum of the random variables is equal to $M$. Equivalently one can think
of an ensemble the configurations of which are defined as masses $\underline
m$ whose sum is $M$, with the weights of the configurations being $%
\prod_{i=1}^L f(m_i)$. The question of condensation then reduces to the
study of the statistical properties of the largest of the $L$ random
variables.

Let us define the moments of $f(m)$ as 
\begin{equation}
\mu _k=\int_0^\infty \ensuremath{\mathrm{d}}m\,m^kf(m)\;.  \label{mu1}
\end{equation}
If the mean $\mu _1$ exists (e.g., (\ref{fm1})) and is such that $L\mu _1>M$%
, we expect the ensemble to be dominated by configurations where $m_i=%
\mathcal{O}(1)$ $\forall i$. However, if $L\mu _1<M$ then we expect the
ensemble to be dominated by configurations where $L-1$ random variables are $%
\mathcal{O}(1)$ and one is $O(M)$. Thus, for $f(m)\sim m^{-\gamma }$ and $%
\gamma >2$ we expect condensation at a critical density $\rho _c=\mu _1$.
This recovers the results given in section~\ref{sec:gce}.

We can also obtain the asymptotic behaviour of the partition function (\ref
{Z}) by noting that the large random variable could be any of the $L$
possible ones and that its probability $\sim A(M-M_c)^{-\gamma }$. Therefore
we expect in the condensed phase 
\begin{equation}
Z(M,L)\sim AL/(M-M_c)^\gamma\;.  \label{Zsc}
\end{equation}
It turns out that this simple argument gives the correct asymptotic behavior
to be derived in section~\ref{sec:cangen} (see Eqs. (\ref{pfcon2}), (\ref
{pfntail})).

Let us now consider the fluctuations of the condensate in the case $\gamma
>2 $. Since we have the constraint of total mass $M$ we can equally well
consider the fluctuations of the total fluid mass which is essentially the
sum of $L-1$ random variables drawn from $f(m)$: 
\begin{equation}
M_f=\sum_{i=1}^{L-1}m_i\;.
\end{equation}
Now if $\gamma >3$, then $\mu _2$ is finite and 
\begin{equation}
\Delta \equiv \sqrt{\mu _2-\mu _1^2}  \label{Delta}
\end{equation}
is the well defined width of $f\left( m\right) $. The standard results of
the Central Limit Theorem applies to this fluid component, with 
\begin{eqnarray}
\Delta M_f &\equiv &\left[ \langle M_f^2\rangle -\langle M_f\rangle
^2\right] ^{1/2} \\
&\simeq &\left[ (\mu _2-\mu _1^2)L\right] ^{1/2}\quad \mbox{for}\quad \gamma
>3
\end{eqnarray}
By constraint, the fluctuation of the condensate will be controlled by $%
\Delta \sqrt{L}$.

However if $2<\gamma\leq 3$, then $\mu_2$ does not exist. Instead, $\Delta
M_f \simeq \sum_i m_i^2$ will be dominated by the largest of the $L-1$
random variables in the fluid. Therefore 
\begin{equation}
\Delta M_f = \mathcal{O}(L^{\frac{1}{\gamma-1}}) \quad \mbox{for} \quad 2 <
\gamma \leq 3\;.
\end{equation}
In this case the fluctuations of the condensate are large $\Delta M_f \gg
O(L^{1/2})$ and we refer to this as an anomalous condensate.

%{\bf remove ?}
%Now let us consider again the case $1<\gamma \leq 2$, where the mean $\mu_1$
%does not exist. In this case, by a usual extremal statistics argument, we
%expect the largest of the $L$ random variables drawn from $f(m)$ to be $%
%\mathcal{O}(L^y)$ where $y=1/(\gamma -1)$. Therefore the average total mass
%would be dominated by the largest and also would be $\mathcal{O}(L^y)$. When 
%$M \gg L^y$, which corresponds to superextensive global mass, one of the
%random variables has to be much larger than expected. It is tempting to
%think of this as condensation, but as we shall see, although a single large
%mass does appear there is no phase transition. We shall refer to this as
%pseudocondensation.

In fact, using the sums of random variables interpretation we obtain very
quickly the scaling behaviour of our main results. In the following sections
we will obtain more precise and detailed results.

\section{Nature of the Condensate: Summary of Results}

\label{sec:nat} 
\begin{figure}[tbph]
\epsfxsize=8cm \centerline{\epsfbox{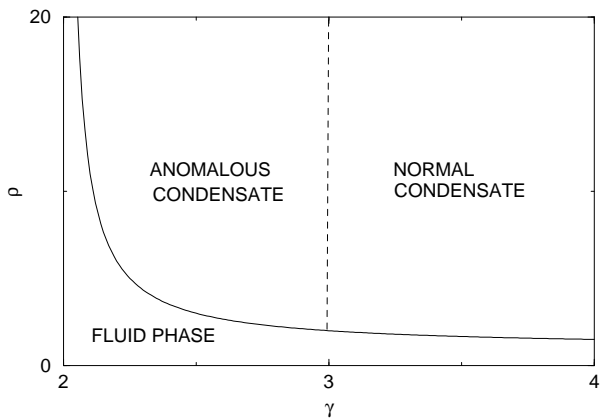}}
\caption{Schematic phase diagram in the $\rho $--$\gamma $ plane.}
\label{fig:phd}
\end{figure}

The GCE analysis of section~\ref{sec:gce} correctly predicts
the criterion for condensation and even the critical density $\rho
_c=\int_0^\infty mf(m)\ensuremath{\mathrm{d}}m$ (whenever condensation
occurs), but provides little insight into the condensed phase itself where $%
\rho >\rho _c$. In this section we explore the condensed phase in detail by
staying within the framework of a `canonical ensemble' and analyzing the
single site mass distribution 
\begin{equation}
p(m)\equiv \int \ensuremath{\mathrm{d}}m_2....\ensuremath{\mathrm{d}}%
m_L\,P(m,m_2,\cdots ,m_L)\delta \left( \sum_{j=2}^Lm_j+m-M\right) ,
\label{pm0}
\end{equation}
in a finite system of size $L$ and total mass $M=\rho L$. Note that $p(m)$
depends on $L$, though we have suppressed its $L$ dependence just for
notational simplicity. Using Eqs. (\ref{prob}) and (\ref{Z}), we have 
\begin{equation}
p(m)=f(m)\frac{Z(M-m,L-1)}{Z(M,L)}.  \label{pm1}
\end{equation}
The rest of this section is devoted to the analysis of $p(m)$ in (\ref{pm1})
with $f(m)$ given by (\ref{fm1}). We have thus two parameters $\gamma $ and $%
\rho $. Our goal is to show how the condensation manifests itself in
different behaviors of $p(m)$ in different regions of the $(\rho -\gamma )$
plane. We will show that for $\gamma >2$, there is a critical curve $\rho
_c(\gamma )$ in the $(\rho -\gamma )$ plane that separates a fluid phase
(for $\rho <\rho _c(\gamma )$) from a condensed phase (for $\rho >\rho
_c(\gamma )$). In the fluid phase the mass distribution decays exponentially
for large $m$, $p(m)\sim \exp [-m/m^{*}]$ where the characteristic mass $%
m^{*}$ increases with increasing density and diverges as the density
approaches its critical value $\rho _c$ from below. At $\rho =\rho _c$ the
distribution decays as a power law, $p(m)\sim m^{-\gamma }$ for large $m$.
For $\rho >\rho _c$, the distribution, in addition to the power law decaying
part, develops an additional bump, representing the condensate, centred
around the ``excess'' mass: 
\begin{equation}
M_{ex}\equiv M-\rho _c\,L.  \label{Mexcess}
\end{equation}
Furthermore, by our analysis within the canonical ensemble, we show that
even inside the condensed phase ($\rho >\rho _c(\gamma )$), there are two
types of behaviors of the condensate depending on the value of $\gamma $.
For $2<\gamma <3$, the condensate is characterized by anomalous non-gaussian
fluctuations whereas for $\gamma >3$, the condensate has gaussian
fluctuations. This leads to a rich phase diagram in the $(\rho -\gamma )$
plane, a schematic picture of which is presented in Fig. (\ref{fig:phd}). 

To proceed, we take the Laplace transform of (\ref{Z}): 
\begin{equation}
\int_0^\infty Z(M,L)e^{-sM}\ensuremath{\mathrm{d}} M=\left[ g(s)\right] ^L\;,
\label{lt1}
\end{equation}
where 
\begin{equation}
g(s)=\int_0^\infty f(m)e^{-sm}\ensuremath{\mathrm{d}} m\;.
\end{equation}
The main challenge is to invert (\ref{lt1}) for a given $f(m)$ to compute $%
Z(M,L)$ then exploit its behavior via (\ref{pm1}) to analyse the single site
mass distribution $p(m)$. It turns out that in certain cases, to be
discussed in section \ref{sec:exact}, one can invert (\ref{lt1}) exactly.
However, for general $f(m)$ we rely on an asymptotic analysis to be
presented in section~\ref{sec:cangen}. Before presenting the details of
these calculations we give a summary of our main results for $p(m)$

\subsection{Summary of Results}

\label{sec:sum}

\noindent \textbf{Fluid phase} $\rho<\rho_c$: In this case one finds (see 
\ref{pmfluid2},\ref{mch1}) 
\begin{equation}
p(m)\sim f(m) \, e^{-m/m^*}\quad\mbox{for}\quad 1\ll m \ll M
\end{equation}
where the characteristic mass $m^*$ diverges $\rho$ approaches $\rho_c$ from
below as $(\rho-\rho_c)^{-1}$ for $\gamma >3$ and $(\rho-\rho_c)^{-1/(%
\gamma-2)}$ for $2<\gamma <3$.\\

%{\bf so far  we have not included large $M$ cut-off in fluid distribution}

\noindent \textbf{Condensed Phase} $\rho >\rho _c$: In this case one finds 
\begin{eqnarray}
p(m) &\simeq &f(m)\quad \mbox{for}\quad 1\ll m\ll O(L) \\
p(m) &\simeq &f(m)\frac 1{(1-x)^\gamma }\quad \mbox{for}\quad m=xM_{ex}\quad %
\mbox{where}\quad 0<x<1 \\
p(m) &\sim &p_{\mathrm{cond}}(m)\quad \mbox{for}\quad m\sim M_{ex}
\end{eqnarray}
Here $p_{\mathrm{cond}}$ is the piece of $p(m)$ which describes the
condensate: Centred on the excess $M_{ex}$ and with integral being equal to $%
1/L$, it takes on two distinct forms (see \ref{pcondm1},\ref{pcpeak})
according to whether $\gamma <3$ or $\gamma >3$. For $2<\gamma <3$ 
\begin{equation}
p_{\mathrm{cond}}(m)\simeq L^{-\gamma /(\gamma -1)}\,V_\gamma \left[ \frac{%
m-M_{ex}}{L^{1/{\gamma -1}}}\right] ,
\end{equation}
where $V_\gamma (z)$ is given exactly by Eq. (\ref{scf1}) with asymptotic
forms in Eqs. (\ref{vzn}-\ref{vzp}). Thus the shape of the condensate bump
is non-gaussian for $2<\gamma <3$ and we refer to this as an `anomalous'
condensate. On the other hand, for $\gamma >3$ 
\begin{equation}
p_{\mathrm{cond}}(m)\simeq \frac 1{\sqrt{2\pi \Delta ^2L^3}%
}\,e^{-(m-M_{ex})^2/{2\Delta ^2L}}\quad \mathrm{for}\,\,|m-M_{ex}|\ll
O(L^{2/3}).
\end{equation}
i.e. $p_{\mathrm{cond}}(m)$ is gaussian on the scale $|m-M_{ex}|\ll
O(L^{2/3})$, but, far to the left of the peak, $p(m)$ decays as a power law
(see \ref{pcntail}).\\ 

\noindent \textbf{Critical density} $\rho=\rho_c$: In this case one finds
(see \ref{pmc2},\ref{pmc3}) that 
\begin{eqnarray}
p(m)&\propto& f(m) V_{\gamma}\left(m/L^{1/(\gamma-1)}\right), \quad\mbox{for}%
\quad 2<\gamma<3 \\
p(m)&\propto& f(m) \, e^{-m^2/{2\Delta^2 L}}\quad \gamma>3 .  \label{m2}
\end{eqnarray}
where the scaling function $V_{\gamma}(z)$ is given by (\ref{scf1}). Thus at
criticality $p(m)$ decays as a power law $m^{-\gamma}$ for large $m$ which
is cut-off by a finite size scaling function and the cut-off mass scales as 
\begin{eqnarray}
m_{\mathrm{cut-off}} &\sim & L^{1/(\gamma-1)} \quad\quad \mathrm{for}\,\,\,
2<\gamma<3  \label{m1} \\
&\sim & L^{1/2} \quad\quad \mathrm{for}\,\,\, \gamma>3 .
\end{eqnarray}

\section{Condensation in the Canonical Ensemble: Exactly Solvable Cases}

\label{sec:exact} Before proceeding to derive the results of section \ref
{sec:sum} in the general case, it is useful to work out cases where both $%
Z(M,L)$ and $p(m)$ can be obtained in closed form. We consider two examples
here. In the first example there is a genuine condensation transition,
whereas in the second one has only a pseudocondensation.

\textbf{Example I:} Making the choice 
\begin{equation}
f(m) = \frac{2}{\sqrt{\pi}} \frac{1}{m^{5/2}}\, e^{-1/m},  \label{w1}
\end{equation}
for all $m$, yields the large $m$ behavior (\ref{fm1}) with $\gamma= 5/2$
and $A = 2/\sqrt{\pi}$. To calculate the Laplace transform $g(s)$ of $(\ref
{w1})$ we make use of the following identity, \textbf{3.471}.9 from
Gradshtyn and Ryzhik\cite{GR}, 
\begin{equation}
\int_0^{\infty}x^{\nu-1} e^{-\beta/x -\gamma x} \ensuremath{\mathrm{d}} x =
2 \left(\frac{\beta}{\gamma}\right)^{\nu/2} K_{\nu}\left(2\sqrt{\beta \gamma}%
\right),  \label{i1}
\end{equation}
where $K_{\nu}(x)$ is the modified Bessel function. Choosing $\beta=1$, $%
\gamma=s$, $\nu=-3/2$ and using the fact that $K_{-\nu}(x)=K_{\nu}(x)$ one
gets 
\begin{equation}
\int_0^{\infty} x^{-5/2} e^{-1/x-sx} \ensuremath{\mathrm{d}} x= 2 s^{3/4}
K_{3/2}\left(2\sqrt{s}\right).  \label{i2}
\end{equation}
Next, we use the expression 
\begin{equation}
K_{3/2}(z) = \sqrt{\frac{\pi}{2z}} \frac{(1+z)}{z} e^{-z}.  \label{i3}
\end{equation}
Substituting this on the r.h.s of Eq. (\ref{i2}) and simplifying we obtain 
\begin{equation}
g(s)= \frac{2}{\sqrt{\pi}} \int_0^{\infty} m^{-5/2} e^{-1/m-sm} %
\ensuremath{\mathrm{d}} m = (1+2\sqrt{s}) e^{-2\sqrt{s}}.  \label{i4}
\end{equation}
We then substitute this result on the r.h.s. of Eq. (\ref{lt1}) and expand
the r.h.s to obtain 
\begin{equation}
\int_0^{\infty} Z(M,L) e^{-s M} \ensuremath{\mathrm{d}} M = \sum_{k=0}^L {%
\binom{L }{k}} {(4s)}^{k/2} e^{-2L \sqrt{s}}.  \label{lt2}
\end{equation}

To proceed we need to invert the Laplace transform in Eq. (\ref{lt2}). For
this, we require the inverse Laplace transform, $\mathcal{L}%
_s^{-1}\left[s^{k/2}e^{-a \sqrt{s}}\right]$ with $a=2L$. This is done by
again using the identity in Eq. (\ref{i1}). Using $\nu=-1/2$, $\beta=a^2/4$, 
$\gamma=s$ and 
\begin{equation}
K_{1/2}(z)= \sqrt{\pi/{2z}}\, e^{-z}  \label{K1/2}
\end{equation}
we get 
\begin{equation}
\int_{0}^{\infty} x^{-3/2} e^{-a^2/{4x}} e^{-sx} \ensuremath{\mathrm{d}} x = 
\frac{2\sqrt{\pi}}{a} e^{-a \sqrt{s}}.  \label{inv1}
\end{equation}
This provides us with the identity, 
\begin{equation}
\mathcal{L}_s^{-1}\left[ e^{-a \sqrt{s}}\right] = \frac{a\, e^{-a^2/4x}}{2%
\sqrt{\pi} x^{3/2}} = -\frac{1}{\sqrt{\pi x}} \frac{\partial}{\partial a}
\left[ e^{-a^2/{4x}}\right]\;,  \label{inv2}
\end{equation}
where $x$ is the argument of the inverse Laplace transform. Next we
differentiate both sides of Eq. (\ref{inv2}) $k$ times with respect to $a$.
This gives, 
\begin{eqnarray}
\mathcal{L}_s^{-1}\left[ s^{k/2} e^{-a \sqrt{s}}\right]&=& \frac{(-1)^{k-1}}{%
\sqrt{\pi x}} {\left(\frac{\partial}{\partial a}\right)}^{k+1} \left[e^{-a^2/%
{4x}}\right]  \nonumber \\
&=& \frac{(4x)^{-(k+1)/2}}{\sqrt{\pi x}} e^{-a^2/{4x}} H_{k+1}\left(\frac{a}{%
\sqrt{4x}}\right),  \label{inv3}
\end{eqnarray}
where $H_k(x)$ is the Hermite polynomial of degree $k$ and argument $x$ and
we have used the Rodrigues' representation 
\begin{equation}
H_k(x) = (-1)^k e^{x^2} \frac{\ensuremath{\mathrm{d}}^k}{\ensuremath{%
\mathrm{d}} x^k} e^{-x^2}\;.
\end{equation}
Using the result (\ref{inv3}), we then invert the Laplace transform in Eq. (%
\ref{lt2}) to obtain 
\begin{equation}
Z(M,L) = \frac{1}{2\sqrt{\pi}} \frac{e^{-L^2/M}}{M} \sum_{k=0}^L {\binom{L}{k%
}} M^{-k/2} H_{k+1}\left(\frac{L}{\sqrt M}\right).  \label{pf2}
\end{equation}

We still need to perform the sum in Eq. (\ref{pf2}). To do this, we use the
following identity which is proved in the appendix-A, 
\begin{equation}
\sum_{k=0}^L {\binom{L }{k}} b^{-k}\, H_k(x) = b^{-L}\, H_{L} (x+ b/2),
\label{h1}
\end{equation}
where $b$ is any constant. Now, consider the sum 
\begin{eqnarray}
\sum_{k=0}^{L} {\binom{L }{k}} \frac{H_{k+1}(x)}{b^k} &=& \sum_{k=0}^{L} {%
\binom{L }{k}} b^{-k}\, \left[2x\, H_k (x) - \frac{dH_k(x)}{dx}\right] 
\nonumber \\
&=& 2x\, b^{-L}\, H_L(x+b/2) - \frac{d}{dx}\left[ b^{-L}\, H_L(x+b/2)\right]
\nonumber \\
&=& 2x\, b^{-L}\, H_{L}(x+b/2) - 2L\, b^{-L}\, H_{L-1} (x+b/2).  \label{h2}
\end{eqnarray}
In going from 1st to the second line of Eq. (\ref{h2}), we have used the
well known identity, $H_{k+1}(x)= 2x H_k(x) - dH_k/dx$, in going from 2nd to
the 3rd line we have used the identity in Eq. (\ref{h1}) and in going from
3rd to the 4th line we have used another well known identity $dH_k(x)/dx =
2k H_{k-1}(x)$. Substituting this result on the r.h.s. of Eq. (\ref{pf2})
and simplifying we obtain our final formula, valid for all $M$ and $L$, 
\begin{equation}
Z(M,L)= \frac{L}{\sqrt{\pi}} M^{-(L+3)/2}\,e^{-L^2/M}\,\left[H_L\left(\frac{%
2L+M}{2\sqrt{M}} \right)- \sqrt{M}\, H_{L-1}\left(\frac{2L+M}{2\sqrt{M}}%
\right)\right].  \label{pf3}
\end{equation}
As a check for consistency, one can easily verify that for $L=1$, using $%
H_0(x)=1$ and $H_1(x)=2x$, the formula in Eq. (\ref{pf3}) reduces to 
\begin{equation}
Z(M, 1) = \frac{2}{\sqrt{\pi}} \frac{1}{M^{5/2}}\, e^{-1/M} = f(M),
\label{z1}
\end{equation}
as it should be for $L=1$ from the definition in Eq. (\ref{Z}).

\begin{figure}[htbp]
\epsfxsize=8cm \centerline{\epsfbox{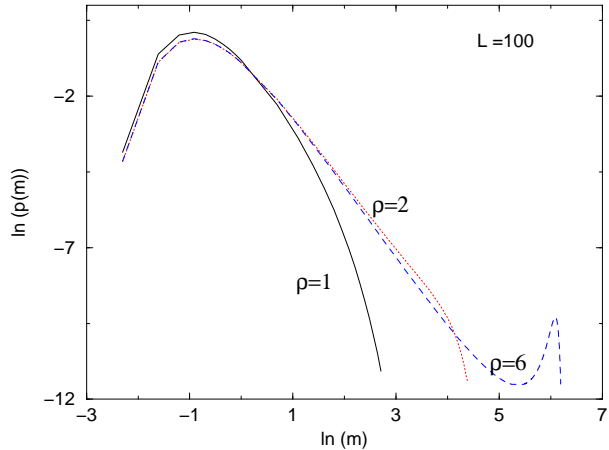}}
\caption{The distribution $p(m)$ vs. $m$ for the exactly solvable case,
plotted using \textit{Mathematica} for $L=100$ and $\rho=1$ (subcritical), $%
\rho=\rho_c=2$ (critical) and $\rho=6$ (supercritical). The condensate shows
up as an additional bump near the tail of $p(m)$ in the supercritical case.}
\label{fig:escpm}
\end{figure}

Since $f(m)\sim m^{-5/2}$ for large $m$, we expect from the GCE analysis in
section-3 that in this case there will be a condensation transition at the
critical density, $\rho_c= \rho(\mu=0)$, where $\rho(\mu)$ is given by Eq. (%
\ref{cp1}). Now, $\rho(\mu=0)$ can be obtained from the Laplace transform: $%
\rho(\mu=0)=-g^{\prime}(0)$, where $g^{\prime}(s)=\ensuremath{\mathrm{d}} g/%
\ensuremath{\mathrm{d}} s$. Using Eq. (\ref{i4}) we get 
\begin{equation}
\rho_c= 2 \;.  \label{critden}
\end{equation}
To see how the condensation transition manifests itself in the single site
mass distribution function $p(m)$, we calculate $p(m)$ explicitly by
substituting in Eq. (\ref{pm1}) the expression for $f(m)$ from Eq. (\ref{w1}%
) and that of $Z(M,L)$ from Eq. (\ref{pf3}). Using Mathematica, we plot this
explicit expression of $p(m)$ for $L=100$ for three densities, $\rho=1$
(subcritical), $\rho=\rho_c=2$ (critical) and $\rho=6$ (supercritical) in
Fig. \ref{fig:escpm}. The transition from the fluid phase ($\rho<\rho_c=2$)
to the condensed phase ($\rho>\rho_c=2$) is clearly seen: the condensate
appears as an additional bump in $p(m)$ near its tail for the supercritical
case $\rho=6$. 

The exact solution detailed above is instructive. It confirms the grand
canonical prediction that the signature of the condensation transition is
manifest in the behaviour of the single site mass distribution $p(m)$ near
its tail, as evident in Fig. (\ref{fig:escpm}). The asymptotic decay of $p(m)
$ for large $m$ is different for $\rho<\rho_c$, $\rho=\rho_c$ and $%
\rho>\rho_c$. For $\rho<\rho_c$, $p(m)$ decays exponentially for large $m$.
For the critical case $\rho=\rho_c$, $p(m)$ has a power law tail: $p(m)\sim
m^{-5/2}$ for large $m$. As $\rho$ increases beyond $\rho_c$, the critical
power law part of $p(m)$ does not change, but the extra mass $(\rho-\rho_c)L$
accumulates on the condensate which shows up as an additional bump beyond
the power law tail of $p(m)$. Thus for $\rho>\rho_c$, $p(m)$ has two parts:
a power law decaying part followed by a bump, indicating that the condensate
coexists with a background that behaves as a critical fluid. Even though
this exact solution was obtained only for a specific $f(m)$ (\ref{w1}), it
confirms the generic picture of section~\ref{sec:sum} for the behaviour in
the tail of $p(m)$ at subcritical, critical and the supercritical densities
respectively. Our goal in the next section is to analyse the large $m$
behaviour of $p(m)$ for these three cases for a general $f(m)$ allowing a
condensation transition, as in Eq. (\ref{fm1}).

\begin{figure}[htbp]
\epsfxsize=8cm \centerline{\epsfbox{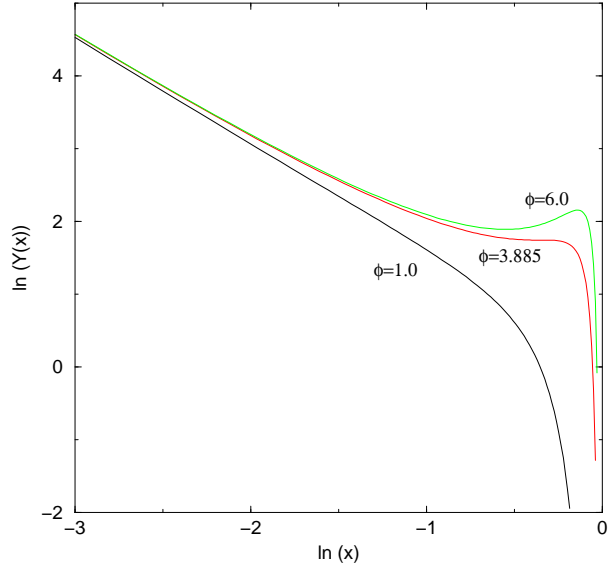}}
\caption{The scaling function $Y(x)$ vs. $x$ in Eq. (\ref{yx1}), for $\phi=1$%
, $\phi=\phi_c=3.88562$ and $\phi=6$. The pseudocondensate shows up as a
bump near the tail of $Y(x)$ for $\phi>\phi_c$.}
\label{fig:pscond}
\end{figure}

\textbf{Example II:} It is also instructive to solve exactly a case where $%
f(m)$ decays more slowly than $1/m^2$ and condensation does not occur. This
leads us to our second exactly solvable example where we find that even
though there is no genuine condensation, a pseudocondensate appears
nevertheless.

We consider 
\begin{equation}
f(m) = \frac{1}{\sqrt{\pi}} \frac{1}{m^{3/2}}\, e^{-1/m},
\end{equation}
for which, using (\ref{K1/2}), one finds 
\begin{equation}
g(s) = e^{-2s^{1/2}}\;.
\end{equation}
Inverting (\ref{lt1}), using (\ref{inv2}), yields 
\begin{equation}
Z(M,L) = \frac{1}{\pi^{1/2}} \frac{L}{M^{3/2}} e^{-L^2/M}
\end{equation}
and from (\ref{pm1}) we find 
\begin{equation}
p(m) = \frac{1}{\pi^{1/2}} \frac{L-1}{L} \left(\frac{M}{m(M-m)}\right)^{3/2}
\exp( -(L-1)^2/(M-m) +L^2/M -1/m )
\end{equation}

A natural scaling regime emerges when $M= \phi L^2, m= xM$. Note that in
this regime the density is superextensive.  Taking the limit of $L$ large we
obtain $p(x\, M)= {\pi^{-1/2}}\, M^{-3/2}\, Y(x)$ where 
\begin{equation}
Y(x) = \frac{1}{[x(1-x)]^{3/2}} \exp[-x/(\phi(1-x))]\;.  \label{yx1}
\end{equation}
A bump in the tail of this distribution emerges when there are two turning
points of $Y(x)$ at positive $x$. This occurs when $\phi > \phi_c$ where $%
\phi_c =2 + 4 \sqrt{2}/3=3.88562\dots $ (see Fig.\ \ref{fig:pscond}). As we
shall discuss later in section~{\ref{sec:pc} this bump is not a true
condensate but rather corresponds to what we shall refer to as a
`pseudocondensate'. }

\section{Condensation in the Canonical Ensemble: General Case}

\label{sec:cangen} 
\begin{figure}[htbp]
\epsfxsize=10cm \centerline{\epsfbox{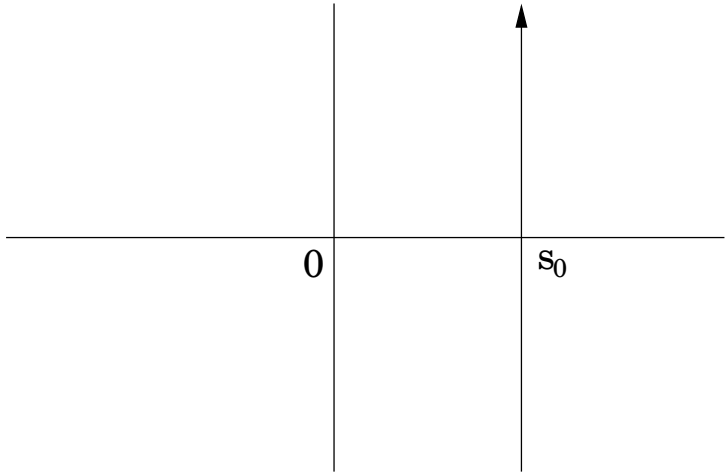}}
\caption{The contour parallel to the imaginary axis passing through the
saddle point $s_0$ in the complex $s$ plane.}
\label{fig:saddle}
\end{figure}

We now proceed to analyze the case of a general $f(m)$
characterized by a power law tail with exponent $\gamma>2$ as in Eq. (\ref
{fm1}). As already noted the Laplace transform 
\begin{equation}
g(s)\equiv \int_0^{\infty} f(m) e^{-s\,m}\, \ensuremath{\mathrm{d}} m
\label{ltfm}
\end{equation}
plays a crucial role in our analysis. We start by formally inverting the
Laplace transform in Eq. (\ref{lt1}) using the Bromwich formula, 
\begin{equation}
Z(M,L)=\int_{s_0-i\infty }^{s_0+i\infty }\frac{\ensuremath{\mathrm{d}} s}{%
2\pi i}\exp \left[ L\left( \ln g(s)+\rho s\right) \right]  \label{brom1}
\end{equation}
where we have used $M=\rho L$. Similarly, one has 
\begin{equation}
Z(M-m,L)=\int_{s_0-i\infty }^{s_0+i\infty }\frac{\ensuremath{\mathrm{d}} s}{%
2\pi i}\exp \left[ L\left( \ln g(s)+\rho s\right)-s\, m \right].
\label{brom2}
\end{equation}
In Eqs. (\ref{brom1}) and (\ref{brom2}), the contour in the complex $s$
plane parallels the imaginary axis with its real part, $s_0$, to the right
of all singularities of the integrand. Since $f(m<0)\equiv 0$, the integrand
is analytic in the right half plane. Therefore, $s_0$ can assume any
non-negative value. Meanwhile, for $f$ given by (\ref{fm1}), $s=0$ is a
branch point singularity. As we shall see, in the subcritical case $%
\rho<\rho_c$ there exists a saddle point at positive $s$ and $s_0$ can be
chosen to be this saddle point (see Fig. (\ref{fig:saddle})). One can then
evaluate the Bromwich integral in Eq. (\ref{brom1}) by the saddle point
method for large $L$. As the density $\rho$ approaches its critical value $%
\rho_c$ from below, the saddle point $s_0$ moves towards $0$. Thus, in the
critical ($\rho=\rho_c$) and in the supercritical ($\rho>\rho_c$), one can
no longer evaluate the Bromwich integral by the saddle point method and an
alternative approach must be used. 

Let us first evaluate the integral in Eq. (\ref{brom1}) in the limit $L\to
\infty $ by the saddle point method, assuming it exists. It is convenient to
define 
\begin{equation}
h(s)\equiv \rho s+\ln g(s).  \label{hs1}
\end{equation}
Then the saddle point equation, $h^{\prime }(s_0)=0$, is 
\begin{equation}
\rho =-g^{\prime }(s_0)/g(s_0),  \label{rhocon}
\end{equation}
which is precisely equivalent to GCE approach in Eq. (\ref{cp1}) on
identifying $s_0=\mu$. Thus the saddle point evaluation works provided there
exists a solution $s_0>0$ of Eq. (\ref{rhocon}). As clarified in the GCE
analysis, a solution exists when $\rho< \rho_c$ where 
\begin{equation}
\rho_c = -g^{\prime}(0)/g(0) = -g^{\prime}(0).  \label{rhoc1}
\end{equation}
We have assumed that $f(m)$ is normalized to unity so that $g(0)=1$. Also,
when the saddle point $s_0>0$ exists, one simply has 
\begin{equation}
Z(M,L)\simeq \frac{\exp (Lh(s_0))}{\sqrt{2\pi Lh^{\prime \prime }(s_0)}}.
\label{sada}
\end{equation}
Similarly when $m \ll O(L)$, from Eq. (\ref{brom2}) one gets $Z(M-m,L)\simeq
Z(M,L) e^{-s_0 m}$. Substituting in Eq. (\ref{pm1}) we get, for $\rho <\rho
_c$ and $1 \ll m\ll O(L)$, 
\begin{equation}
p(m)\simeq f(m)\mathrm{e}^{-s_0 m},  \label{pmfluid}
\end{equation}
thus recovering the GCE result upon identifying $s_0=\mu$. This shows that
for $\rho<\rho_c$, the canonical and the grand canonical approach are
equivalent and the mass distribution $p(m)$ decays exponentially for large $%
m $ with a characteristic mass $m^*=1/s_0$.

As $\rho$ approaches the critical value $\rho_c$ from below, $s_0$
approaches $0$ and hence the characteristic mass $m^*$ diverges. To see how
the asymptotic behaviour of $p(m)$ changes as one increases $\rho$ through
its critical value, we will now focus only in the vicinity of the critical
point. Near the critical point, i.e, when $|\rho-\rho_c|$ is small, the most
important contribution to the integral in Eq. (\ref{brom1}) comes from the
small $s$ region. Hence, to obtain the leading behaviour for large $L$ it is
sufficient to consider only the small $s$ behavior of $g(s)$ in Eq. (\ref
{brom1}). For $f(m)$ in Eq. (\ref{fm1}) with a noninteger $\gamma>2 $, one
can expand, quite generally, the Laplace transform $g(s)$ in Eq. (\ref{ltfm}%
) for small $s$, as 
\begin{equation}
g(s)=\sum_{k=0}^{n-1}(-1)^k\frac{\mu _k}{k!}s^k+bs^{\gamma -1}+\ldots
\label{gsexp}
\end{equation}
Here $n=\mbox{int}[\gamma ]$, $\mu _k$ is the $k^{th}$ moment of $f(m)$
(which exists for $k<n$). We assume that $f(m)$ is normalized to unity,
i.e., $\mu_0=1$. Note also that from Eq. (\ref{rhoc1}) it follows that 
\begin{equation}
\mu_1= \rho_c .  \label{mom1}
\end{equation}
The term $b s^{\gamma-1}$ in Eq. (\ref{gsexp}) is the leading singular term.
For $f(m)$ in Eq. (\ref{fm1}) with a noninteger $\gamma$, one can show (see
appendix-B) that 
\begin{equation}
b = A\, \Gamma(1-\gamma),  \label{bamp}
\end{equation}
a relation that we will use later. Note that the expansion in Eq. (\ref
{gsexp}) is valid only for noninteger $\gamma$. For integer $\gamma$, one
can carry out a similar analysis and it is easy to show that the leading
singular term has a logarithmic correction: 
\begin{equation}
g(s)=\sum_{k=0}^{\gamma-2}(-1)^k\frac{\mu _k}{k!}s^k+cs^{\gamma -1}\ln s
+\ldots\quad\mbox{for}\quad \gamma = 2,3,\ldots
\end{equation}
where 
\begin{equation}
c= \frac{(-1)^\gamma A}{(\gamma-1)!}
\end{equation}

Substituting the small $s$ expansion of $g(s)$ from Eq. (\ref{gsexp}) in Eq.
(\ref{hs1}) and keeping only the two leading order terms explicitly one gets 
\begin{eqnarray}
h(s) &=&(\rho -\rho _c)s+bs^{\gamma -1}+\dots \quad \quad \mathrm{for}%
\,\,\,2<\gamma <3  \nonumber \\
&=&(\rho -\rho _c)s-As^2\log (s)/2+\dots \quad \quad \mathrm{for}%
\,\,\,\gamma =3  \nonumber \\
&=&(\rho -\rho _c)s+\Delta ^2s^2/2+\dots +bs^{\gamma -1}+\dots \quad \quad 
\mathrm{for}\,\,\,\gamma >3  \label{las1}
\end{eqnarray}
where $\Delta $ is defined in Eq. (\ref{Delta}). The special role of $\gamma
=3$ is now clear. The next-to-leading term in $h(s)$ is $s^{\gamma -1}$ for $%
2<\gamma <3$ and $s^2$ for $\gamma >3$.

Below we substitute the small $s$ expansion of $h(s)$ from Eq. (\ref{las1})
in the integral in Eq. (\ref{brom1}) and analyse its leading asymptotic
behaviour for large $L$ for the three cases $\rho<\rho_c$, $\rho>\rho_c$ and 
$\rho=\rho_c$ separately. Since we are using the small $s$ expansion, this
analysis is valid only in the vicinity of the critical point, i.e., when $%
|\rho-\rho_c|$ is small.

\subsection{Subcritical Case: $\rho<\rho_c$}

For $\rho<\rho_c$, it follows from Eq. (\ref{las1}) that $h(s)$ has a saddle
at $s_0>0$. This is equivalent to saying that Eq. (\ref{rhocon}) has a
solution which, to leading order in $(\rho_c-\rho)$, is given by 
\begin{eqnarray}
s_0 &\simeq& \left[\frac{(\rho _c-\rho )}{b(\gamma -1)}\right] ^{1/(\gamma
-2)}\quad\quad \mathrm{for}\,\,\, 2< \gamma<3  \nonumber \\
&\simeq & - (\rho_c-\rho)/{A\log (\rho_c-\rho)}\quad\quad \mathrm{for}\,\,\,
\gamma=3  \nonumber \\
&\simeq & (\rho _c-\rho )/\Delta ^2 \quad\quad \mathrm{for}\,\,\, \gamma>3.
\label{s_01}
\end{eqnarray}

Consequently, it follows from Eq. (\ref{pm1}), as well as from Eq. (\ref
{pmfluid}), that the single site mass distribution $p(m)$ behaves for $1\ll
m \ll O(L)$ as 
\begin{equation}
p(m)\sim \frac{A}{m^{\gamma}}\, e^{-s_0 m},  \label{pmfluid2}
\end{equation}
where $s_0$ is given by Eq. (\ref{s_01}) for small $(\rho_c-\rho)$. Thus the
characteristic mass $m^*=1/s_0$ diverges as a power law as $\rho$ approaches 
$\rho_c$ from below, 
\begin{eqnarray}
m^* &\sim & (\rho_c-\rho)^{-1/(\gamma-2)} \quad\quad \mathrm{for}\,\,\, 2<
\gamma<3  \nonumber \\
& \sim & -\log(\rho_c-\rho)/(\rho_c-\rho) \quad\quad \mathrm{for}\,\,\,
\gamma=3  \nonumber \\
& \sim & (\rho_c-\rho)^{-1} \quad\quad \mathrm{for}\,\,\, \gamma>3.
\label{mch1}
\end{eqnarray}

The partition function in Eq. (\ref{sada}), which we will require later, may
also be calculated and behaves asymptotically as 
\begin{eqnarray}
Z(M=\rho L, L) &\sim & \exp\left[-L\, B\,
(\rho_c-\rho)^{(\gamma-1)/(\gamma-2)}\right] \quad\quad \mathrm{for}\,\,\,
2< \gamma<3  \nonumber \\
&\sim & \exp\left[L(\rho_c-\rho)^2/{2 A \log (\rho_c-\rho)}\right]
\quad\quad \mathrm{for}\,\,\, \gamma=3  \nonumber \\
& \sim & \exp\left[-L {(\rho_c-\rho)^2}/{2\Delta^2}\right] \quad\quad 
\mathrm{for}\,\,\, \gamma>3,  \label{pfsp1}
\end{eqnarray}
where $B= (\gamma-2)\, b^{-1/(\gamma-2)}\,
(\gamma-1)^{-(\gamma-1)/(\gamma-2)}$.

\subsection{Supercritical case: $\rho>\rho_c$}

For the \emph{supercritical} regime ($\rho >\rho _c$), there is no solution
to (\ref{rhocon}) on the positive real axis and more care is needed to find
the asymptotic form of $Z(M,L)$. One can see this clearly in the vicinity of
the critical point where one can use the small $s$ expansion of $h(s)$ in
Eq. (\ref{las1}): evidently for $\rho>\rho_c$, $h(s)$ does not have a
minimum at any positive $s_0$. In the $L\to \infty$ limit and with $%
(\rho-\rho_c)$ small, one can, however, develop a scaling analysis by
identifying the different scaling regimes and calculating the corresponding
scaling forms for $Z(M,L)$ and hence that of $p(m)$. The main result of this
subsection is to obtain an exact asymptotic formula describing the shape of
the condensate bump.

Substituting Eq. (\ref{gsexp}) in Eqs. (\ref{brom1}) and (\ref{brom2}), we
get 
\begin{eqnarray}
Z(M,L)&\simeq & W_{L}(\rho_c-\rho)  \label{spf1} \\
Z(M-m,L) & \simeq & W_{L}\left((m-M_{ex})/L\right),  \label{spf2}
\end{eqnarray}
where $M_{ex}\equiv (\rho -\rho _c)L$ is the excess mass and 
\begin{equation}
W_L(y)=\int_{-i\infty }^{i\infty }\frac{\ensuremath{\mathrm{d}} s}{2\pi i}%
\exp \left[ L(-ys+ C(s))\right].  \label{wy1}
\end{equation}
In Eq. (\ref{wy1}) the function $C(s)$ represents the nonanalytic correction
to the linear term in $h(s)$ in Eq. (\ref{las1}), i.e., 
\begin{eqnarray}
C(s)&=& b s^{\gamma-1} \quad\quad \mathrm{for}\,\,\, 2< \gamma<3  \nonumber
\\
&=& - A s^2 \log(s)/2\quad\quad \mathrm{for}\,\,\, \gamma=3  \nonumber \\
&=& \Delta^2 s^2/2 + \dots + bs^{\gamma-1}\quad\quad \mathrm{for}\,\,\,
\gamma>3  \label{cs1}
\end{eqnarray}
By including the first nonanalytic term in $C(s)$ we will obtain the leading
large $L$ asymptotic behaviour.

Knowing the function $W_L(y)$ and using Eqs. (\ref{spf1}) and (\ref{spf2}),
one can rewrite the mass distribution in Eq. (\ref{pm1}) as 
\begin{equation}
p(m)\simeq f(m)\frac{W_L\left( (m-M_{ex})/L\right) }{W_L\left( \rho
_c-\rho\right) },  \label{pm2}
\end{equation}
All crucial information about the condensate `bump' around $%
m=M_{ex}=(\rho-\rho_c)L$ is thus encoded in the asymptotic behavior of $%
W_L(y)$ defined in Eq. (\ref{wy1}). Below we analyze the asymptotics of $%
W_L(y)$ and hence that of $p(m)$ in the two cases $2<\gamma<3$ and $\gamma>3$
separately.

\vspace{0.5cm}

\textbf{Case-I \,\, $2<\gamma<3$ : } In this case, we substitute $C(s)=
bs^{\gamma-1}$ in Eq. (\ref{wy1}) and rescale $s\to L^{-1/(\gamma-1)}s$ to
rewrite Eq. (\ref{wy1}) in the scaling form 
\begin{equation}
W_L(y) = L^{-1/(\gamma-1)} V_{\gamma}\left[
L^{(\gamma-2)/(\gamma-1)}y\right],  \label{wy2}
\end{equation}
where the scaling function is 
\begin{equation}
V_{\gamma}(z)=\int_{-i\infty }^{i\infty }\frac{\ensuremath{\mathrm{d}} s}{%
2\pi i}e^{-zs+bs^{\gamma -1}}.  \label{scf1}
\end{equation}
We were not able to perform the complex integral in Eq. (\ref{scf1}) in
closed form. However, its asymptotic behaviors can be worked out, the
details of which are relegated to appendix-C. We find, 
\begin{eqnarray}
V_\gamma (z) &\simeq &A\,|z|^{-\gamma }\quad \mathrm{as}\,\,z\to -\infty
\label{vzn} \\
&=& c_0 \quad \mathrm{at}\,\, z=0  \label{vz0} \\
&\simeq & c_1\, z^{(3-\gamma)/{2(\gamma-2)}}\, e^{-c_2
z^{(\gamma-1)/(\gamma-2)}} \quad \mathrm{as}\,\, z\to \infty  \label{vzp}
\end{eqnarray}
where $A$ is the amplitude in Eq. (\ref{fm1}), $b=A\, \Gamma(1-\gamma) $ (as
in Eq. (\ref{bamp})) and the constants $c_0$, $c_1$ and $c_2$ are given as 
\begin{eqnarray}
c_0 &=&
b^{-1/(\gamma-1)}/[(\gamma-1)\Gamma\left((\gamma-2)/(\gamma-1)\right)]
\label{cons0} \\
c_1 &=& \left[2\pi (\gamma-2) (b(\gamma-1))^{1/(\gamma-2)}\right]^{-1/2}
\label{cons1} \\
c_2 & =& (\gamma-2)/{(\gamma-1)(b(\gamma-1))^{1/(\gamma-2)}} .  \label{cons2}
\end{eqnarray}

\begin{figure}[htbp]
\epsfxsize=8cm \centerline{\epsfbox{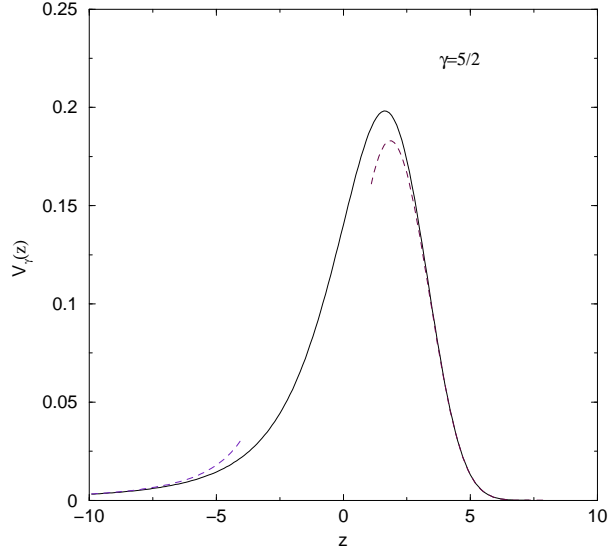}}
\caption{Plot of the scaling function $V_{\gamma}(z)$ for $\gamma=5/2$
obtained by numerically integrating the integral in Eq. (\ref{scf2}). The
dashed lines show the asymptotic tails obtained analytically in Eqs. (\ref
{vzn}) and (\ref{vzp}).}
\label{fig:vz}
\end{figure}

Evidently the scaling function $V_{\gamma}(z)$ is highly asymmetric, has an
algebraic decay $|z|^{-\gamma}$ as $z\to -\infty$ and decays extremely fast
(faster than a gaussian) as $z\to \infty$. To plot the full function $%
V_{\gamma}(z)$ in Eq. (\ref{scf1}), it is useful to first transform the
integration in the complex $s$ plane in Eq. (\ref{scf1}) to a real integral.
This can be achieved by making a change of variable $s=e^{\pm i\pi/2}y$
respectively for the upper and the lower half of the imaginary axis in the
integral in Eq. (\ref{scf1}) and then simplifying. One obtains 
\begin{equation}
V_{\gamma}(z)=\frac{1}{\pi} \int_0^{\infty} \ensuremath{\mathrm{d}} y\,
e^{-c_3 y^{\gamma-1}} \cos\left[b\cos(\pi\gamma/2)y^{\gamma-1}+yz\right],
\label{scf2}
\end{equation}
where $c_3= -b\sin(\pi\gamma/2)>0 $ for $2<\gamma<3$. The real integral in
Eq. (\ref{scf2}) can be easily evaluated numerically. A plot of this
function for $\gamma=5/2$, where the integral in Eq. (\ref{scf2}) was
performed numerically using Mathematica, is shown in Fig. (\ref{fig:vz}).
The dashed lines at the two tails show the agreement with the asymptotic
analytical forms in Eqs. (\ref{vzn}) and (\ref{vzp}). 

Armed with knowledge of the asymptotic behaviors of $W_L(y)$ we are now
ready to calculate the mass distribution $p(m)$ from Eq. (\ref{pm2}). Let us
first evaluate the denominator in Eq. (\ref{pm2}) using the scaling form of $%
W_L(y)$ in Eq. (\ref{wy2}), 
\begin{equation}
Z(M,L)\simeq W_L(\rho _c-\rho )=L^{-1/(\gamma -1)}V_\gamma \left[
-L^{(\gamma -2)/(\gamma -1)}(\rho -\rho _c)\right] .  \label{pfcon1}
\end{equation}
Since $\rho >\rho _c$, the argument of the scaling function in Eq. (\ref
{pfcon1}) is a large negative number as $L\to \infty $, provided $\rho -\rho
_c=M_{ex}/L>0$ is kept fixed. One can then use the asymptotic tail of $%
V_\gamma (z)$ as given in Eq. (\ref{vzn}) and we get, in the regime where $%
M_{ex}\sim O(L)$, 
\begin{equation}
Z(M,L)\simeq W_L(\rho _c-\rho )\simeq \frac A{L^{\gamma -1}\,(\rho -\rho
_c)^\gamma }=\frac{A\,L}{M_{ex}{}^\gamma }.  \label{pfcon2}
\end{equation}
Collecting all these results together in Eq. (\ref{pm2}) one then finds 
\begin{equation}
p(m)\simeq \frac{f(m)}A\,(\rho -\rho _c)^\gamma \,L^{\gamma (\gamma
-2)/(\gamma -1)}\,V_\gamma \left[ \frac{m-M_{ex}}{L^{1/{\gamma -1}}}\right] .
\label{pm3}
\end{equation}
Note that the result in Eq. (\ref{pm3}) is valid for all $m$, provided the
total mass $M$ and and the system size $L$ are both large with $M/L=\rho $
fixed. Now one can identify various limits of (\ref{pm3}). Using (\ref{vzn})
one finds 
\begin{eqnarray}
p(m) &\simeq &f(m)\quad \mbox{for}\quad 1\ll m\ll O(L) \\
p(m) &\simeq &f(m)\frac 1{(1-x)^\gamma }\quad \mbox{for}\quad m=xM_{ex}\quad %
\mbox{where}\quad 0<x<1
\end{eqnarray}
Thus for $m\ll O(L)$, $p(m)$ is a pure power law but when $m$ becomes
extensive $p(m)$ begins to deviate from $f(m)$.

Focusing now only near the condensate where $m\sim M_{ex}$, one obtains
using $f(m)\sim Am^{-\gamma }$ the asymptotic form of $p(m)\simeq p_{\mathrm{%
cond}}(m)$ near the condensate bump, 
\begin{equation}
p_{\mathrm{cond}}(m)=L^{-\gamma /(\gamma -1)}\,V_\gamma \left[ \frac{m-M_{ex}%
}{L^{1/{\gamma -1}}}\right] ,  \label{pcondm1}
\end{equation}
where $V_\gamma (z)$ is given exactly by Eq. (\ref{scf1}) with asymptotic
forms in Eqs. (\ref{vzn}-\ref{vzp}) and a full shape as shown in Fig. (\ref
{fig:vz}). Thus the shape of the condensate bump, as shown in Fig. (\ref
{fig:vz}) is highly asymmetric and non-gaussian for $2<\gamma <3$ and we
refer to this as the `anomalous' condensate. Also note that, setting $\xi
=(m-M_{ex})L^{-1/{\gamma -1}}$, 
\begin{equation}
\int_{-\infty }^\infty \ensuremath{\mathrm{d}}m\,p_{\mathrm{cond}%
}(m)=L^{-1}\int_{-\infty }^\infty \ensuremath{\mathrm{d}}\xi \,V_\gamma (\xi
)
\end{equation}
Thus the area under $p_{\mathrm{cond}}(m)$ is $O(1/L)$ which corresponds to
precisely one condensate in the system of size $L$.

\vspace{0.5cm}

\textbf{Case-II $\gamma> 3$ : } In this case, using the appropriate form for 
$C(s)$ from Eq. (\ref{cs1}) in Eq. (\ref{wy1}) we get 
\begin{equation}
W_L(y)=\int_{-i\infty }^{i\infty }\frac{\ensuremath{\mathrm{d}} s}{2\pi i}%
\exp \left[ L(-ys + \Delta^2s^2/2 +\dots + b s^{\gamma-1}\right].
\label{wy3}
\end{equation}
We will see that $W_L(y)$ has different behaviors for $y>0$ and $y<0$. Let
us first consider the case $y>0$, where one can again perform a saddle point
analysis as in the previous subsection and one gets, to leading order in $L$
and for $y\sqrt{L}\sim O(1)$ 
\begin{equation}
W_L(y) \simeq \frac{1}{\sqrt{2\,\pi\, \Delta^2\, L}}e^{-y^2 L/{2\Delta^2}}.
\label{wy3p}
\end{equation}
Thus the function $W_L(y)$ has a gaussian decay for positive but small $y$.
A careful analysis shows that the gaussian form with the scaling variable $y%
\sqrt{L}$ in Eq. (\ref{wy3p}) remains valid not just over the range $y\sqrt{L%
}\sim O(1)$, but over a wider range: at least up to $y\sim O(L^{-1/3})$. For
large $y>0$, one can still do a saddle point analysis `a la Eq. (\ref{sada})
and $W_L(y)$ will have a non-gaussian tail for large, positive $y$.

We next turn to $y=-|y|<0$. In this case, there is no saddle point and we
have to do the integration along the imaginary axis passing through the
origin. Again, let us first consider the regime where $|y|\sqrt{L} \sim O(1)$%
. In this case, we rescale $s\to s/\sqrt{L}$ in Eq. (\ref{wy3}) which then
becomes 
\begin{equation}
W_L(y) = \frac{1}{\sqrt{L}}\,\int_{-i\infty}^{i \infty}\frac{%
\ensuremath{\mathrm{d}} s}{2\pi i}\exp \left[-(y\sqrt{L}) s + \Delta^2 s^2/2
+\dots + b L^{(3-\gamma)/2} s^{\gamma-1}\right].  \label{wy3l}
\end{equation}
Keeping $|y|\sqrt{L}=z$ fixed and taking the $L\to \infty$ limit, one can
drop all the subleading terms and the resulting integral is a gaussian one
which can be simply performed to give for $|y|\sim O(1/\sqrt{L})$, 
\begin{equation}
W_L(y) \simeq \frac{1}{\sqrt{2\,\pi\, \Delta^2\, L}}e^{-y^2 L/{2\Delta^2}},
\label{wy3n}
\end{equation}
the same result as for $y\sim O(L^{-1/2})>0$ in Eq. (\ref{wy3p}). Thus,
within the range $|y|\sim O(L^{-1/3})$, the function $W_L(y)$ is symmetric
and gaussian. On the other hand, far to the left of the origin where $|y|
\gg O(L^{-1/3})$, this analysis breaks down. Let us illustrate the case $|y|
\sim O(1)$. In that regime of $|y|$, we again start from Eq. (\ref{wy3}) but
this time rescale $s\to s /L$ which gives 
\begin{eqnarray}
W_L(y)& =& \frac{1}{L} \int_{-i\infty}^{i\infty} \frac{\ensuremath{%
\mathrm{d}} s}{2\pi i}\exp \left[ |y| s + \Delta^2 s^2/{2L}+\dots+ b
L^{2-\gamma} s^{{\gamma-1}} \right]  \label{wy3n1} \\
&\simeq & \frac{1}{L} \int_{-i\infty}^{i\infty} \frac{\ensuremath{\mathrm{d}}
s}{2\pi i}\, \exp[|y|\, s]\, \left[1+ \frac{\Delta^2\,s^2}{2L} +\dots +b 
\frac{s^{\gamma-1}}{L^{\gamma-2}}\right],  \label{wy3n2}
\end{eqnarray}
where in going from first to the second line, we have expanded the
exponential for large $L$, keeping $|y|$ fixed. Now, the integral in Eq. (%
\ref{wy3n2}) can be done term by term. The first term gives a delta function 
$\delta(|y|)$ which is $0$ since $|y|>0$. Similarly the second term, which
is the second derivative of the delta function with respect to $|y|$, is
also zero. In fact, all the analytic terms containing integer powers of $s$
will be similarly equal to $0$, except the last singular term which has $%
s^{\gamma-1}$. Thus one gets for $|y|\sim O(1)$ with $L\to \infty$, 
\begin{equation}
W_L(y) \simeq \frac{b}{L^{\gamma-1}}\, \int_{-i\infty}^{i\infty} \frac{%
\ensuremath{\mathrm{d}} s}{2\pi i}\, \exp[|y|\, s]\, s^{\gamma-1} = \frac{b}{%
|y|^{\gamma}\, L^{\gamma-1}}\, \int_{-i\infty}^{i\infty} \frac{%
\ensuremath{\mathrm{d}} s}{2\pi i}\, e^{s}\,s^{\gamma-1}.  \label{wy3n3}
\end{equation}
The latter integral is done in appendix-C (see Eq. (\ref{int1})). Using this
result in Eq. (\ref{wy3n3}) one obtains 
\begin{equation}
W_L(y) \simeq \frac{A}{|y|^{\gamma}\, L^{\gamma-1}}\quad \mbox{for}\quad
|y|\sim O(1)  \label{wy3n4}
\end{equation}
Although here we consider $|y|\sim O(1)$ it is easy to show the result also
holds for $|y| \sim O(L^{-\alpha})$ where $\alpha <1/2$.

Thus, the large $L$ behavior of the function $W_L(y)$ for $\gamma>3$ can be
summarized as follows, 
\begin{eqnarray}
W_L(y) & \simeq & \frac{1}{\sqrt{2\,\pi\, \Delta^2\, L}}\, e^{-y^2 L/{%
2\Delta^2}} \quad\quad \mathrm{for}\,\,\, |y|\ll O(L^{-1/3})  \label{wygpeak}
\\
&\simeq & \frac{A}{|y|^{\gamma}\, L^{\gamma-1}} \quad\quad \mathrm{for}%
\,\,\, -y=|y| \sim O(1)  \label{wyntail}
\end{eqnarray}
The function $W_L(y)$ is thus a gaussian near its peak at $y=0$ and then as
non-gaussian tails far away from the peak. On the negative side, this tail
is algebraic and decays as $|y|^{-\gamma}$.

We now compute the asymptotic behavior of the mass distribution $p(m)$ in
Eq. (\ref{pm2}) for $\gamma >3$. First, we calculate the partition function
in Eq. (\ref{spf1}). Substituting the results in Eqs. (\ref{wygpeak}) and (%
\ref{wyntail}) in Eq. (\ref{spf1}) we find 
\begin{eqnarray}
Z(M,L) &\simeq &W_L\left( -(\rho -\rho _c)\right)   \nonumber \\
&\simeq &\frac{e^{-M_{ex}{}^2/{2\Delta ^2L}}}{\sqrt{2\,\pi \,\Delta ^2\,L}}%
\,\quad \,\,\mathrm{for}\,\,|M_{ex}|\sim O(L^{2/3})  \label{pfgpeak} \\
&\simeq &\frac{A\,L}{M_{ex}{}^\gamma }\quad \quad \mathrm{for}\,\,M_{ex}\sim
O(L).  \label{pfntail}
\end{eqnarray}
Comparing Eqs. (\ref{pfntail}) and (\ref{pfcon2}) one sees that for $M_{ex}>0
$ (i.e., $M>\rho _cL$) and in the regime where it is $O(L)$, the partition
function $Z(M,L)$ has the same asymptotic form for all $\gamma >2$. Using
the results in Eqs. (\ref{wygpeak}) and (\ref{wyntail}) one can similarly
evaluate the numerator in Eq. (\ref{pm2}). Collecting these results, we find
that 
\begin{equation}
p(m)\simeq f(m)\frac{M_{ex}{}^\gamma }{AL}\,W_L\left( \frac{m-M_{ex}}%
L\right) .  \label{pm4}
\end{equation}
The the asymptotic form of $p(m)=p_{\mathrm{cond}}(m)$ near the condensate
bump has a gaussian peak centered at $m=M_{ex}=(\rho -\rho _c)\,L$ with its
width scaling as $O(\sqrt{L})$, but has a non-gaussian tail far away from
the peak. More precisely, one gets the following behavior near the peak: 
\begin{equation}
p_{\mathrm{cond}}(m)\simeq \frac 1{\sqrt{2\pi \Delta ^2L^3}%
}\,e^{-(m-M_{ex})^2/{2\Delta ^2L}}\quad \mathrm{for}\,\,(m-M_{ex})\sim
O(L^{2/3}).  \label{pcpeak}
\end{equation}
On the other hand, for $m$ far less than the peak value of $M_{ex}$, we have 
\begin{equation}
p_{\mathrm{cond}}(m)\simeq f(m)\,(1-m/M_{ex})^{-\gamma }\quad \mathrm{for}%
\,\,M_{ex}-m\sim O(L).  \label{pcntail}
\end{equation}
The integral of $p_{\mathrm{cond}}$ gives $1/L+O(L^{1-\gamma })$ where the
dominant contribution is from the gaussian peak. Again this implies a single
condensate in the system.

\subsection{Critical Case: $\rho=\rho_c$}

In the critical case $\rho=\rho_c$, it follows from Eq. (\ref{pm2}) 
\begin{equation}
p(m) \simeq f(m) \frac{W_L\left(m/L\right)}{W_L(0)}.  \label{pmc1}
\end{equation}
We now consider the two cases $2<\gamma<3$ and $\gamma>3$ separately.

\vspace{0.5cm}

\textbf{Case-I $2<\gamma<3$:} In this case, using Eq. (\ref{wy2}) we get 
\begin{equation}
p(m)\simeq f(m)\frac{V_{\gamma}\left(m/L^{1/(\gamma-1)}\right)}{V_{\gamma}(0)%
}=\frac{f(m)}{c_0}\,V_{\gamma}\left(m/L^{1/(\gamma-1)}\right),  \label{pmc2}
\end{equation}
where the constant $c_0$ is given in Eq. (\ref{cons0}) and the asymptotic
behavior of the function $V_{\gamma}(z)$ can be read off Eqs. (\ref{vz0})
and (\ref{vzp}). The Eq. (\ref{pmc2}) thus describes the finite size scaling
function associated with the distribution $p(m)$: it decays as a power law $%
p(m)\simeq f(m)\sim A m^{-\gamma}$ for large $m$ and then is cut off by the
finite size of the sample and the cut-off mass scales as $m_{\mathrm{cut-off}%
}\sim L^{1/(\gamma-1)}$.

\vspace{0.5cm}

\textbf{Case-II $\gamma>3$:} In this case, using Eq. (\ref{wy3p}) in Eq. (%
\ref{pmc1}) we get 
\begin{equation}
p(m)\simeq f(m) \, e^{-m^2/{2\Delta^2 L}}\quad\quad \mathrm{for} \,\,\,
m\sim O(L^{2/3}).  \label{pmc3}
\end{equation}
Thus, in this case, the cut-off function is Gaussian up to $m\sim O(L^{2/3})$%
, even though the cut off mass scales as $m_{\mathrm{cut-off}} \sim L^{1/2}$
for all $\gamma>3$.

\vspace{0.5cm}

\section{Other cases}

\label{sec:oc} So far we have only considered $f(m)$ with asymptotic form (%
\ref{fm1}). In this section we extend our results to two other choices of $%
f(m)$.

\subsection{Stretched exponential case}

First we briefly discuss the case 
\begin{equation}
f(m) \simeq A\exp (-c m^\alpha) \quad \mbox{where}\quad \alpha <1\;.
\end{equation}
This stretched exponential decay clearly fulfils the criterion for
condensation of section~\ref{sec:gce}. Noting that all moments $\mu_k$ of $f$
exist, one finds that the subcritical partition function in the vicinity of
the critical point behaves asymptotically as 
\begin{equation}
Z(M=\rho L, L) \sim \exp\left[-L {(\rho_c-\rho)^2}/{2\Delta^2}\right]
\end{equation}
and the subcritical $p(m)$ behaves as 
\begin{equation}
p(m) \sim A \exp(-c m^\alpha -s_0 m),
\end{equation}
where $s_0=(\rho_c-\rho)/\Delta^2$. Thus in the subcritical regime, there
are two `mass' scales. The first one is a natural scale of $O(1)$ that
characterises the stretched exponential decay of $f(m)$ itself and it
remains of $O(1)$ even at the critical point. The second mass scale $%
1/s_0\sim (\rho_c-\rho)^{-1}$, that emerges out of collective behavior,
however diverges as the critical point is approached from below. Thus,
unlike the power law $f(m)$ discussed in the previous section where at the
critical point $p(m)$ becomes scale free, for the stretched exponential case
only one out of the two scales diverges at the critical point. In this
sense, the condensation transition associated with the stretched exponential
case is somewhat different from the usual scenario of a second order phase
transition.

In the condensed phase $\rho>\rho_c$, one can make an analysis qualitatively
similar to the power-law case with $\gamma>3$, though the details are
somewhat different. Omitting details, we find that for $\rho>\rho_c$ the
condensate bump is gaussian near its peak as expected 
\begin{equation}
p_{\mathrm{cond}}(m) \simeq \frac{1}{\sqrt{2\pi \Delta^2 L^3}}\,
e^{-(m-M_{ex})^2/{2\Delta^2 L}} \quad \mathrm{for}\,\, (m-M_{ex})\sim O(%
\sqrt{L}).
\end{equation}
On the other hand, far to the left of the peak where $M_{ex}-m \sim O(L)$ we
have 
\begin{equation}
p_{\mathrm{cond}}(m) \simeq f(m)\, \exp( c \left[ M_{ex}^{\alpha}
-|m-M_{ex}|^{\alpha}\right])  \label{setail}
\end{equation}
%{\bf Actually I have not shown this explicitly just assume it will be the case..
%needs to be checked}

\subsection{Case $1<\gamma <2$}

\label{sec:pc} As noted in section \ref{sec:rv}, in the case where 
\begin{equation}
f(m) \simeq A m^{-\gamma} \quad 1<\gamma <2  \label{fg12}
\end{equation}
we do not expect true condensation. However, as we shall now analyse, an
interesting phenomenon may occur at suitably chosen superextensive densities
where a `pseudocondensate' bump emerges.

For (\ref{fg12}) the corresponding small $s$ expansion of $g(s)$ reads 
\begin{equation}
g(s) = \mu_0 + b s^{\gamma-1} + \cdots
\end{equation}
where $b = A \Gamma(1-\gamma)$ and we take as usual $\mu_0 =1$. Now since $%
b<0$ for $1< \gamma < 2$ we \emph{always} find a solution $s_0 >0$ to the
saddle point equation (\ref{rhocon}) which when small is 
\begin{equation}
s_0 \simeq \left[ \frac{-b(\gamma-1)L}{M}\right]^{1/(2-\gamma)}
\end{equation}
and the saddle point expression for $Z(M,L)$ reads 
\begin{equation}
Z(M,L) \simeq \left(2\pi (\gamma-2)(\gamma-1)b L
s_0^{\gamma-3}\right)^{-1/2} \exp \left\{ -\frac{2-\gamma}{\gamma-1}\left( 
\frac{-b (\gamma-1) L }{M^{\gamma-1}}\right)^{1/(2-\gamma)} \right\}\;.
\label{Zg12}
\end{equation}

Let us consider the regime where 
\begin{equation}
M=\phi L^{1/(\gamma -1)}\qquad m=xM  \label{pcscale}
\end{equation}
where $\phi $ and $x$ are fixed as $L\to \infty $. Using (\ref{pm1}) we find 
\begin{equation}
p(x)\simeq \frac A{M^\gamma }x^{-\gamma }(1-x)^{-(3-\gamma )/{2(2-\gamma )}%
}\exp \left[ -C\left( (1-x)^{-(\gamma -1)/(2-\gamma )}-1\right) \right] 
\end{equation}
where 
\begin{equation}
C=(2-\gamma )(-b)^{1/(2-\gamma )}\left( \frac{\gamma -1}\phi \right)
^{(\gamma -1)/(2-\gamma )}
\end{equation}
As $\phi $ is increased a bump in the tail of $p(x)$ emerges corresponding
to two real turning points of $p(x)$. However we argue that this does not
correspond to a true condensate for the following reasons. Firstly, as $M$
increases, $Z(M,L)$ will follow the same expression (\ref{Zg12}) thus there
can be no true phase transition since $Z(M,L)$ is analytic. Secondly, there
is no diverging length scale in $p(m)$. Thirdly, the bump in $p(x)$ is broad
i.e. extends over a finite range of $x$, unlike a true condensate bump which
would be narrow. Therefore we call this bump a `pseudocondensate' as it is
really an extension of the fluid phase, rather than a condensate coexisting
with the fluid.

To understand why (\ref{pcscale}) is the natural scale for $M$, let us
consider the sum of $L$ random variables each drawn from a distribution (\ref
{fg12}). We expect the largest of the $L$ random variables drawn from $f(m)$
to be $\mathcal{O}(L^{1/(\gamma -1)})$ and the total mass to also be $%
\mathcal{O}(L^{1/(\gamma -1)})$. Therefore when $M$ is on the scale (\ref
{pcscale}) the constraint of fixed total mass leads to non-trivial $p(m)$.

%Following is a long winded set of comments, for you rather than the paper.
%To get to the other end, look for another double row of ==='s, ok?

\section{Discussion}

\label{sec:disc} In this work we have presented an analysis of the partition
function (\ref{Z}) of a factorised steady state within the canonical
ensemble. The analysis has revealed the structure of the condensate as
summarised in section~\ref{sec:sum}. The two types of condensate occurring
when $f(m) \sim m^{-\gamma}$ with $2< \gamma < 3$ or $\gamma >3$ correspond
to condensates with anomalous (non-gaussian) and gaussian fluctuations
respectively and each has a distinct scaling function for the shape of the
condensate. In both cases the nonequilibrium phase transition to the
condensed phase is continuous in the sense that the characteristic mass of
the exponential site mass distribution of the fluid phase diverges as the
transition is approached from below. We have also analysed the case of
stretched exponential $f(m)$ where condensation occurs but the transition is
somewhat different: even though there is a diverging scale one gets a
stretched exponential rather than power law distribution $p(m)$ at the
critical point and in the fluid component of the condensed phase. Finally we
have shown that in the case where $f(m) \sim m^{-\gamma}$ where $1 <\gamma
<2 $ a pseudocondensate appears at superextensive critical density, but
there is no phase transition.

Our results may easily be generalised to the case of discrete mass,
exemplified by the ZRP where the occupation of each site is $0,1,2\ldots$
and the total mass $M$ is a positive integer. In this case the relevant
expression for the canonical partition function is 
\begin{equation}
Z(M,L) = \oint \, \frac{dz}{2\pi i} \ z^{-(M+1)}\ \left[ F(z)\right]^L\; ,
\label{Zint}
\end{equation}
where the integral is around a closed contour about the origin in the
complex $z$ plane and 
\begin{equation}
F(z) = \sum_{m=0}^{\infty} z^{m} \ f(m)\;.  \label{Fdef}
\end{equation}
The phase transition occurs when the saddle point of the $z$ integral
reaches the radius of convergence of (\ref{Fdef}). For $f(m)$ of the form (%
\ref{fm1}) the radius of convergence is $z=1$, therefore we let $z=(1-u)$
and expand $F(z)$ for $u$ small as 
\begin{equation}
F(1-u) = \sum_{k=0}^{r-1} \frac{ (- u)^k F^{(k)}}{k!} + b u^{b-1} + \ldots
\label{FB1u}
\end{equation}
where $r$ is the integer part of $\gamma$ and $F^{(k)} = \frac{d^k}{d z^k}%
\left. F(z)\right|_{z=1}$ The second term in (\ref{FB1u}) is the leading
singular part and the coefficient $b$ is again $b= A \Gamma(1-\gamma)$. The
general analysis and results then follow analogously to that of section~\ref
{sec:cangen}.

The structure of the condensate also has implications for the dynamics
within the steady state. In the steady state a site must be randomly
selected to hold the condensate thus the translational symmetry is broken.
However on any finite lattice there will be a timescale $\tau_L$ over which
the condensate dissolves and reforms on another spontaneously selected site.
In systems with symmetry breaking, the `flip time' $\tau_L$ \cite{EFGM} is
of interest. A recent study \cite{GL} has shown that for the ZRP with
asymmetric mass transfer and dynamics which yield $f(m) \sim m^{-\gamma}$
the flip time grows as $\tau_L \sim L^{\gamma}$. This result was found
numerically and also from a simple effective desciption for the condensate
dynamics. That effective description requires a structure for $p(m)$, in
particular a dip to the left of the condensate bump, which is borne out by
the exact results presented here.

Finally we mention that it would be interesting to analyse condensation in
steady states that are not factorised, such as in the chipping model of ~%
\cite{MKB}. Generally, such steady states exhibit stronger correlations than
a factorised one. In such cases a factorised steady state is often used as a
`mean-field' approximation in the sense that some of the correlations in the
true steady state are ignored. Condensation is known to occur in some cases
of non-factorised steady states~\cite{MKB}, on the other hand there are some
examples where the mean field approximation predicts a condensation
transition but in the true steady state there is none~\cite{RK}. It is of
importance to further investigate these issues.\\

\noindent \textbf{Acknowledgments:} MRE thanks LPTMS for hospitality during
a visit when some part of this paper was written and acknowledges partial
support of EPSRC Programme grant GR/S10377/01. RKPZ acknowledges the support
of the US National Science Foundation through DMR-0414122.

\appendix

\section{Proof of an addition theorem for Hermite Polynomials (\ref{h1})}

In this appendix, we prove the following identity, 
\begin{equation}
\sum_{k=0}^L {\binom{L }{k}} b^{-k}\, H_k(x) = b^{-L}\, H_{L} (x+ b/2),
\label{a1}
\end{equation}
The generating function of the Hermite polynomials is well known, 
\begin{equation}
\sum_{k=0}^{\infty} \frac{t^k}{k!} H_k(x)= e^{-t^2 + 2t\,x}.  \label{a2}
\end{equation}
One also has trivially 
\begin{equation}
\sum_{m=0}^{\infty} \frac{ t^{m} b^{m}}{m!} = e^{bt}.  \label{a3}
\end{equation}
Multiplying Eqs. (\ref{a2}) and (\ref{a3}) one gets 
\begin{equation}
\sum_{k,m=0}^{\infty} \frac{ t^{m+k} b^m}{m! k!} H_k(x)=
e^{-t^2+2t(x+b/2)}=\sum_{n=0}^{\infty}\frac{t^{n}}{n!} H_{n}(x+b/2).
\label{a4}
\end{equation}
Matching powers of $t^L$ on both sides of Eq. (\ref{a4}) one arrives at the
result in Eq. (\ref{a1}).

\section{Proof of expansion (\ref{gsexp},\ref{bamp})}

In this appendix we show that for $f(m)$ in Eq. (\ref{fm1}) with a
noninteger $\gamma>2$, the coefficient $b$ in the singular term in the small 
$s$ expansion of the Laplace transform $g(s)$ in Eq. (\ref{gsexp}) is
related simply to the amplitude $A$ of the power law tail of $f(m)$ via 
\begin{equation}
b= A\, \Gamma(1-\gamma).  \label{b1}
\end{equation}

Note that $n$ in Eq. (\ref{gsexp}) is simply $n=\mbox{int}[\gamma ]$.
Differentiating Eq. (\ref{gsexp}) $n$ times with respect to $s$ and using
the definition of $g(s)$ in Eq. (\ref{ltfm}), we get 
\begin{equation}
(-1)^n\, \int_0^{\infty} e^{-s\,m}\, m^n \, f(m)\, \ensuremath{\mathrm{d}}
m= b \frac{\Gamma(\gamma)}{\Gamma(\gamma-n)} s^{\gamma-n-1} +\ldots.
\label{b2}
\end{equation}
Let us denote the l.h.s. of Eq. (\ref{b2}) by $I(s)$. Making a change of
variable $y=sm$ in the integral on the l.h.s of Eq. (\ref{b2}), yields 
\begin{equation}
I(s)= (-1)^n s^{-n-1} \int_0^{\infty} e^{-y} y^n f(y/s) \ensuremath{%
\mathrm{d}} y.  \label{b3}
\end{equation}
Next we take the $s\to 0$ limit in Eq. (\ref{b3}). In that limit, the
leading contribution to the integral will come from the region where the
argument $y/s$ of $f(m)$ is large, so that one can use $f(y/s)\approx A
(s/y)^{\gamma}$. Substituting this expression and performing the resulting
integral, gives, to leading order in $s$, 
\begin{equation}
I(s) = (-1)^n A \Gamma(n+1-\gamma) s^{\gamma-n-1} + \ldots  \label{b4}
\end{equation}
Comparing the leading terms on the l.h.s. and the r.h.s. of Eq. (\ref{b2})
we get 
\begin{equation}
b= \frac{(-1)^n}{\Gamma(\gamma)}\Gamma(n+1-\gamma)\Gamma(\gamma-n)\, A .
\label{b5}
\end{equation}
Simplifying Eq. (\ref{b5}) using the well known identity, 
\begin{equation}
\Gamma(x)\Gamma(1-x)=\pi/{\sin(\pi x)}\;,  \label{Eulide}
\end{equation}
gives the result in Eq. (\ref{b1}).

\section{Asymptotic Expansion of $V_{\gamma}(z)$}

In this appendix we analyse the scaling function $V_{\gamma}(z)$ 
\begin{equation}
V_{\gamma}(z)=\int_{-i\infty }^{i\infty }\frac{\ensuremath{\mathrm{d}} s}{%
2\pi i}e^{-zs+bs^{\gamma -1}}\quad\mbox{where} \quad b= A \Gamma(1-\gamma)\;,
\end{equation}
near its tails $z\to \pm \infty$ for all $2<\gamma<3$ and also evaluate the
function at $z=0$. \vspace{0.5cm}

\textbf{The tail $z\to -\infty$:} We write $z=-|z|$ in Eq. (\ref{scf1}) and
rescale $s\to s/|z|$ to write 
\begin{eqnarray}
V_{\gamma}(z)&=& \frac{1}{|z|}\int_{-i\infty }^{i\infty }\frac{%
\ensuremath{\mathrm{d}} s}{2\pi i}\, e^{s + b\, (s/|z|)^{\gamma-1}} 
\nonumber \\
&=& \frac{1}{|z|}\int_{-i\infty }^{i\infty }\frac{\ensuremath{\mathrm{d}} s}{%
2\pi i}\, e^{s}\left[1+ a\frac{s^{\gamma-1}}{|z|^{\gamma-1}} +\dots\right].
\label{C1}
\end{eqnarray}
In going from the first to the second line in Eq. (\ref{C1}) we have
expanded $\exp[b (s/|z|)^{\gamma-1}]$ in a power series. We now perform the
integration in Eq. (\ref{C1}) term by term. The first term is simply zero
for any nonzero $|z|$. The second term is of order $O(|z|^{-\gamma})$. In
general, the $(n+1)$-th term will scale as $O(|z|^{n(\gamma-1)+1})$. Thus,
for large $|z|$, the leading asymptotic behavior is captured by the second
term and one gets 
\begin{equation}
V_{\gamma}(z) \simeq \frac{b}{|z|^{\gamma}}\int_{-i\infty }^{i\infty }\frac{%
\ensuremath{\mathrm{d}} s}{2\pi i}\, e^{s}\, s^{\gamma-1}.  \label{C2}
\end{equation}
This integral may be performed by wrapping the contour around the branch cut
on the negative real axis i.e. we integrate along $s= e^{\pm i\pi}x$ with $x$
real and positive. Then one finds 
\begin{equation}
\int_{-i\infty }^{i\infty }\frac{\ensuremath{\mathrm{d}} s}{2\pi i}\,
e^{s}\, s^{\gamma-1} = \frac{\sin(\pi \gamma)}{\pi} \Gamma(\gamma)
\label{int1}
\end{equation}
and so as $z\to -\infty$, 
\begin{equation}
V_{\gamma}(z) \simeq \frac{b\,\Gamma(\gamma)\,\sin(\pi \gamma)}{\pi\,
|z|^{\gamma}}= \frac{A}{|z|^{\gamma}}.  \label{C4}
\end{equation}

%\begin{equation}
%V_{\gamma}(z) \simeq \frac{b}{\pi |z|^{\gamma}}\int_0^{\infty} dx\, x^{\gamma-1}\, \sin(x |z| + \pi \gamma/2).
%\label{C3}
%\end{equation}
%The real integral in Eq. (\ref{C3}) can be performed by using the two following identities\cite{GR}
%\begin{eqnarray}
%\int_0^{\infty} dx\, x^{\gamma-1}\, \sin(x) &=& \Gamma(\gamma) \sin(\pi\gamma/2) \label{int1}\\
%\int_0^{\infty} dx\, x^{\gamma-1}\, \cos (x) &=& \Gamma(\gamma) \cos(\pi\gamma/2) \label{int2}.
%\end{eqnarray}

\vspace{0.5cm}

\textbf{The tail $z\to +\infty$:} In this case, the asymptotic behavior of
the integral in Eq. (\ref{scf1}) can be evaluated by the saddle point
method. The action $h_1(s)=-zs+b s^{\gamma-1}$ inside the exponential has a
saddle point on the positive real axis at $s^*=
[z/b(\gamma-1)]^{1/(\gamma-2} $. Since the function is analytic in the plane
between the imaginary axes passing through $s=0$ and $s=s^*$, the contour in
Eq. (\ref{scf1}) can be shifted to $s=s^*$ and then the leading contribution
to this integral for large $z$ will come from the region around the saddle
point at $s=s^*$. Expanding $h_1(s)= h_1(s^*) + (s-s^*)^2
h_1^{\prime\prime}(s^*)/2 +..$ around the saddle and performing the
resulting Gaussian integration, one gets the positive tail of the scaling
function as in Eq. (\ref{vzp}).

\vspace{0.5cm}

\textbf{The value at $z=0$:} We now evaluate $V_{\gamma}(0)$ and show that
its value is given as in Eqs. (\ref{vz0}) and (\ref{cons0}). We need to
perform the integral in Eq. (\ref{scf1}) along the imaginary axis passing
through $s=0$. Rescaling $s\to s b^{-1/(\gamma-1)}$ we get 
\begin{equation}
V_{\gamma}(0)= b^{-1/(\gamma-1)}\int_{-i\infty }^{i\infty }\frac{%
\ensuremath{\mathrm{d}} s}{2\pi i}\,e^{s^{\gamma -1}}  \label{C5}
\end{equation}
To evaluate this integral we bend the contour along the rays $s=\exp^{\pm
i\pi/(\gamma-1)}y$ (where $y$ is real and positive) so that the argument of
the exponential in (\ref{C5}) becomes real and negative. Then one finds 
\begin{eqnarray}
\int_{-i\infty }^{i\infty }\frac{\ensuremath{\mathrm{d}} s}{2\pi i}%
\,e^{s^{\gamma -1}} &=& \frac{\sin (\pi/(\gamma-1))}{\pi} \int_{0 }^{\infty }%
\ensuremath{\mathrm{d}} y\,e^{-x^{\gamma -1}}  \nonumber \\
&=& \frac{1}{\gamma-1}\, \Gamma\left( \frac{\gamma-2}{\gamma-1} \right)
\end{eqnarray}
where we have used $\int_0^\infty \ensuremath{\mathrm{d}} x e^{-x^\alpha} =
\Gamma(1/\alpha)/\alpha$ and the identity (\ref{Eulide}). 
%of the imaginary axis in Eq. (\ref{C5}) and use the identity $e^{ix}=\cos (x) +i \sin(x)$
%to rewrite as a real integral
%\begin{equation}
%V_{\gamma}(0)= \frac{1}{\pi\,(\gamma-1)\, b^{1/(\gamma-1)}}\,\int_0^{\infty} dx\, e^{-a_1 x}\, \cos(a_2 x)\, 
%x^{-(\gamma-2)/(\gamma-1)},
%\label{C6}
%\end{equation}
%where $a_1= -\sin(\pi\gamma/2) >0 $ for $2<\gamma<3$ and $a_2=\cos(\pi\gamma/2)$. The integral
%in Eq. (\ref{C6}) can be performed exactly (see page 490 of Ref. \cite{GR}) and we get
%\begin{equation}
%V_{\gamma}(0)= \frac{1}{\pi\,(\gamma-1)\, b^{1/(\gamma-1)}}\, 
%\frac{\Gamma\left(1/(\gamma-1)\right)}{ {\left(a_1^2+a_2^2\right)}^{1/{2(\gamma-1)}}}\,
%\cos\left(\frac{1}{(\gamma-1)}\, \tan^{-1}(a_2/a_1)\right).
%\label{C7}
%\end{equation}
Finally we have 
\begin{equation}
V_{\gamma}(0)= \frac{1}{b^{1/(\gamma-1)}\, (\gamma-1)\,
\Gamma\left((\gamma-2)/(\gamma-1)\right)}.  \label{C8}
\end{equation}


\vspace*{2em}

\begin{thebibliography}{99}
\bibitem{MRE00}  M.R.~Evans, Braz. J. Phys. \textbf{30}, 42 (2000)

\bibitem{ARAP}  J. Krug and J. Garcia, J. Stat. Phys., \textbf{99} 31
(2000); R. Rajesh and S. N. Majumdar, J. Stat. Phys., \textbf{99} 943 (2000).

\bibitem{traffic}  D Chowdhury, L Santen, A Schadschneider Physics Reports 
\textbf{329}, 199 (2000).

\bibitem{OEC}  O.J.~O'Loan, M.R.~Evans and M.E.~Cates, Phys. Rev. E \textbf{%
58}, 1404 (1998)

\bibitem{KLMST}  Y.~Kafri, E.~Levine, D.~Mukamel, G.M.~Sch\"{u}tz and J.~T{%
\"{o}}r{\"{o}}k, Phys.~Rev.~Lett. \textbf{89}, 035702 (2002)

%Title: A Model for Force Fluctuations in Bead Packs

\bibitem{CLMNW}  S.N. Coppersmith, C.-h. Liu, S. Majumdar, O. Narayan, T.A.
Witten Phys. Rev. E., \textbf{53}, 4673 (1996)

\bibitem{MWL04}  D. van der Meer, K. van der Weele and D. Lohse 2004 J.
Stat. Mech.:Theory and Experiment P04004

\bibitem{Torok}   J. Torok, Physica A \textbf{355} 374-382 (2005).

\bibitem{Jain}  K. Jain, Phys. Rev. E \textbf{72} 017105 (2005).

\bibitem{EH05}  M. R. Evans and T. Hanney, J. Phys.A \textbf{38} R195 (2005).

\bibitem{MEZ05}  S.N. Majumdar, M. R. Evans, R. K. P. Zia, Phys. Rev. Lett., 
\textbf{94} 180601 (2005).

\bibitem{BBJ}  P. Bialas, Z. Burda, and D. Johnston, Nucl. Phys. B \textbf{%
493}, 505 (1997).

\bibitem{Ritort}  F. Ritort 1995 Phys. Rev. Lett. \textbf{75} 1190

\bibitem{DGC}  J. M. Drouffe, C.~Godr\`{e}che and F. Camia 1998 J.~Phys.~A 
\textbf{31} L19

\bibitem{S70}  F.~Spitzer 1970 Adv. Math. \textbf{5} 246

\bibitem{E96}  M. R.~Evans, Europhys. Lett. \textbf{36}, 13 (1996)

\bibitem{KF96}  J.~Krug and P. A. Ferrari 1996 J.~Phys.~A \textbf{29} L465

\bibitem{JMP00}  I. Jeon, P. March and B. Pittel 2000 Ann. Probab. \textbf{28%
} 1162

\bibitem{GSS}  S.~Gro\ss kinsky, G.M Sch\"{u}tz and H.~Spohn, J. Stat. Phys 
\textbf{113}, 389 (2003)

\bibitem{Godreche}  C.~Godr\`{e}che, J. Phys. A: Math. Gen., \textbf{36},
6313 (2003)

\bibitem{AEM}  A. G. Angel, M. R. Evans and D. Mukamel, J. Stat.
Mech.:Theory. Exp. P04001 (2004)

\bibitem{ZS2}  F. Zielen and A. Schadschneider, Phys. Rev. Lett. \textbf{89}%
, 090601 (2002)

\bibitem{EMZ04}  M. R. Evans, S.N. Majumdar, R. K. P. Zia J. Phys. A: Math.
Gen \textbf{37} (2004) L275

\bibitem{MKB}  S.N.~Majumdar, S.~Krishnamurthy and M.~Barma, Phys. Rev.
Lett. \textbf{81}, 3691 (1998) ; J Stat. Phys. \textbf{99 } 1 (2000)

\bibitem{EMZtbp}  M. R. Evans, S.N. Majumdar, R. K. P. Zia \emph{in
preparation}

\bibitem{GL05} R. L. Greenblatt, J. L. Lebowitz
 cond-mat/0506776 
%    Title: Product Measure Steady States of Generalized Zero Range Processes


\bibitem{ZEM04}  R. K. P. Zia, M. R. Evans, S.N. Majumdar, J. Stat. Mech. :
Theor. Exp. (2004) L10001.

\bibitem{RM2}  R. Rajesh and S. N. Majumdar, Phys. Rev. E., \textbf{63},
036114 (2001)

\bibitem{RK}  R. Rajesh and S. Krishnamurthy, Phys. Rev. E. , \textbf{66},
046132 (2002)

\bibitem{LMZ}  E. Levine, D. Mukamel, G. Ziv J. Stat. Mech.: Theor. Exp.
P05001 (2004).

\bibitem{EH03}  M.~R.~Evans and T.~Hanney, J.~Phys.~A \textbf{36}, L441
(2003)

\bibitem{DMS03}  S.N. Dorogovtsev, J.F.F. Mendes, A.N. Samukhin, \textit{%
Nucl. Phys. B} \textbf{666}, 396 (2003).

%Statistical ensemble of scale-free random graphs

\bibitem{BCK01}  Z.~Burda, J.~D.~Correia and A.~Krzywicki, Phys.~Rev.~E 
\textbf{64}, 046118 (2001)

\bibitem{BJJKNPZ02}  Z. Burda, D. Johnston, J. Jurkiewicz, M. Kaminski, M.
A. Novak, G. Papp, I Zahed (2002) \textit{Phys. Rev. E} \textbf{65}, 026102

\bibitem{ELMM}  M. R. Evans, E. Levine, P. K. Mohanty and D. Mukamel 2004
Eur. Phys. J. B \textbf{41} 223

\bibitem{AHR}  P. F. Arndt, T. Heinzel and V. Rittenberg 1998 J. Phys. A:
Math. Gen \textbf{31} L45; 1999 \textit{J. Stat. Phys.} \textbf{97} 1

\bibitem{KSZ}  G. Korniss, B. Schmittmann, R.K.P. Zia Europhys. Lett. 
\textbf{45}, 431 (1999)

\bibitem{RSS}  N. Rajewsky, T. Sasamoto and E. R. Speer 2000 Physica A 
\textbf{279} 123

\bibitem{MSZ}  J.T. Mettetal, B. Schmittmann, R.K.P. Zia Europhys. Lett. 
\textbf{58}, 653 (2002)

\bibitem{KLMT}  Y.~Kafri, E.~Levine, D.~Mukamel and J.~T{\"{o}}r{\"{o}}k
2002 J.~Phys.~A \textbf{35} L459

\bibitem{KLMSW}  Y. Kafri, E. Levine, D. Mukamel, G.M. Sch\"{u}tz, and R. D.
W. Willmann 2003 Phys. Rev. E \textbf{68} 035101

\bibitem{GSZ}  I. T. Georgiev, B. Schmittmann, R. K.P. Zia Phys. Rev. Lett 
\textbf{94}, 115701 (2005)

\bibitem{GR}  I.S. Gradshteyn and I.M. Ryzhik, \emph{Table of Integrals,
series and Products} 6th edition (Academic Press, San Diego, 2000).

\bibitem{EFGM}  M. R. Evans, D. P. Foster, C. Godreche, D. Mukamel, Phys.
Rev. Lett. \textbf{74} 208 (1995).

\bibitem{GL}  C. Godr\`{e}che and J.M. Luck 
% \emph{Dynamics of the condensate in zero-range processes} 
J. Phys. A \textbf{38} 7215-7237 (2005)

%\bibitem{GLEMSS}  C. Godr\`{e}che, J.M. Luck, M.R.Evans, D. Mukamel, S.
%Sandow and E.R.Speer 1995 J. Phys A \textbf{28} 6
\end{thebibliography}
\end{document}